\documentclass[12pt]{article}
\pdfoutput=1 
\usepackage[table]{xcolor}
\usepackage{booktabs}
\usepackage{mathrsfs}
\usepackage{jheppub}
\usepackage{mathtools,leftindex,tensor,mhchem}
\usepackage{amssymb}
\usepackage{dsfont}
\usepackage{amsmath}
\usepackage{bbm}
\usepackage{xcolor}
\usepackage{color, colortbl}
\usepackage{empheq}
\usepackage{tabularx}
\usepackage{amsfonts}
\usepackage{orcidlink}
\usepackage{multicol} \usepackage{multirow}
\usepackage{physics}
\usepackage[most]{tcolorbox}
\usepackage{amsmath}
\usepackage{caption}
\usepackage{subcaption,longtable,stmaryrd}
\usepackage{slashed}
\usepackage{cancel}
\usepackage{multicol}
\usepackage{blindtext}
\usepackage{tikz}
\usepackage{enumitem}
\usepackage[customcolors,shade]{hf-tikz}
\usepackage{graphicx}
\usepackage{cancel}
\DeclareMathAlphabet\mathbfcal{OMS}{cmsy}{b}{n}
\DeclareSymbolFont{usualmathcal}{OMS}{cmsy}{m}{n}
\DeclareSymbolFontAlphabet{\mathcal}{usualmathcal}
\DeclareSymbolFont{rmlargesymbols}{OMX}{mdbch}{m}{n}
\DeclareMathSymbol{\intop}{\mathop}{rmlargesymbols}{82}
\DeclareMathSymbol{\rmointop}{\mathop}{rmlargesymbols}{72}
\definecolor{mygray}{gray}{0.5}

\preprint{\texttt{IFT-UAM/CSIC-26-114}}

\title{Krylov complexity and the growth of the black hole interior in 3D gravity}

\author[a]{Arpan Bhattacharyya,\raisebox{2.5pt}{\scalebox{0.80}{\orcidlink{0000-0002-7933-6441}}}}
\author[a]{Sounak Pal,\raisebox{2.5pt}{\scalebox{0.80}{\orcidlink{0000-0002-2250-0466}}}}
\author[b]{Juan F. Pedraza\raisebox{2.5pt}{\scalebox{0.80}{\,\orcidlink{0000-0002-1426-642X}}}}

\affiliation[a]{\it Indian Institute of Technology, Gandhinagar, Gujarat-382355, India}
\affiliation[b]{\it Instituto de Física Teórica UAM/CSIC, Calle Nicolás Cabrera 13-15, 28049 Madrid, Spain }

\emailAdd{abhattacharyya@iitgn.ac.in}
\emailAdd{palsounak@iitgn.ac.in}
\emailAdd{j.pedraza@csic.es}

\abstract{
We investigate the growth of the black hole interior in three-dimensional
gravity from the boundary theory. For the two-sided BTZ black hole, we
propose a boundary reconstruction of the time dependence of a
codimension-one surface in terms of correlation functions of smeared
operators in the thermofield double state, reproducing the characteristic
late-time linear growth predicted by the complexity-volume proposal. Using the Chern--Simons formulation of three-dimensional gravity, these
nonlocal correlators are represented by bulk Wilson lines with smeared
endpoints, extending the familiar connection between Wilson lines and
codimension-two observables underlying holographic entanglement entropy to
codimension-one observables.
We then ask whether the same geometric growth is captured by Krylov
complexity. For the smeared operators, we extract the Lanczos data from
their correlation functions and find that operator Krylov complexity
reproduces the late-time linear growth of the black hole interior, extending
previous connections between operator growth and bulk geometry to
AdS$_3$. By contrast, the Krylov spread complexity of the thermofield
double state, obtained from the semiclassical gravitational partition
function, does not exhibit the linear growth of the bulk volume within the
regime accessible to our analysis. Our results therefore point to a
distinguished role for operator Krylov complexity in encoding black hole
interior growth beyond two-dimensional gravity, while highlighting a
qualitative distinction between operator and state notions of Krylov
complexity.
}

\makeatletter
\begin{document}

\maketitle


\section{Introduction}

The dynamics of black hole interiors have attracted considerable attention in recent years. Within the AdS/CFT correspondence, one possible boundary characterization of this dynamics is provided by quantum complexity \cite{Susskind:2014rva,Susskind:2014moa,Stanford:2014jda,Brown:2015lvg}. In particular, the Complexity $=$ Volume (CV) proposal relates the complexity of a boundary state to the maximal volume of a codimension-one bulk surface extending through the black hole interior \cite{Stanford:2014jda}. Classically, this volume grows linearly over a parametrically long range of times, whereas non-perturbative quantum effects are expected to eventually halt the growth and produce saturation on time scales of order $e^{S}$.
 
Substantial progress in understanding this phenomenon has been made in two-dimensional gravities, where semiclassical and, in some cases, non-perturbative methods are available \cite{Iliesiu:2021ari,Gautason:2025ryg,Miyaji:2025yvm,Miyaji:2025jxy,Sato:2025uli}. A prototypical example is Jackiw--Teitelboim (JT) gravity, where the codimension-one volume reduces to the length of the Einstein--Rosen bridge (ERB). In the probe approximation, this length can be related to a two-point function of local primary operators inserted on the two boundaries of the eternal black hole. After analytic continuation, the Euclidean correlator gives access to the expectation value of the Lorentzian geodesic length connecting the boundaries. This relation can also be explored non-perturbatively using the random-matrix description of JT gravity. In particular, Ref.~\cite{Iliesiu:2021ari} showed that the non-perturbative contribution to the density-density correlator leads to the saturation of the ERB length, encoding the effects of higher-genus topologies with a fixed boundary component.
 
\begin{figure}[t!]
\centering
\includegraphics[width=0.75\linewidth]{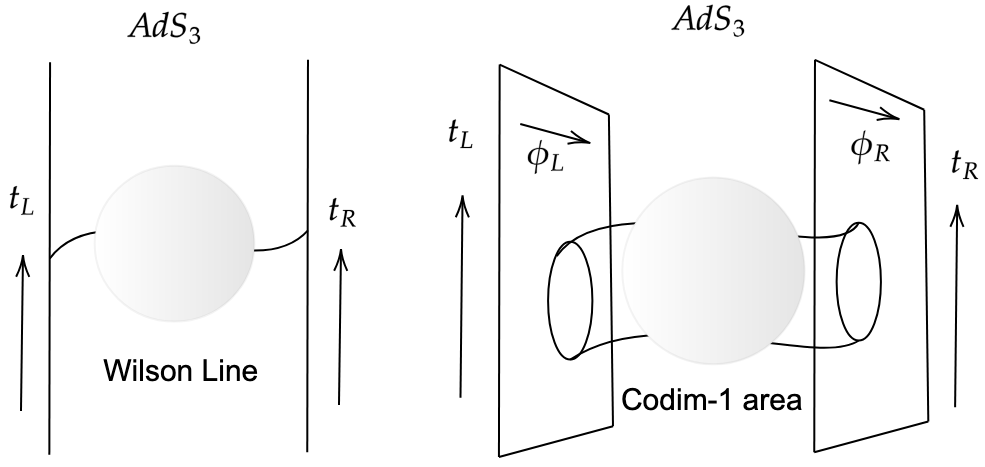}
\caption{Schematic representation of the boundary and bulk constructions. On the left, a Wilson line connecting boundary insertions probes a codimension-two bulk object in AdS$_3$. On the right, we extend this construction by smearing the boundary operators along the periodically identified spatial direction $\phi$. The resulting non-local operators provide the boundary observable used to probe the growth of the codimension-one surface extending through the black hole interior.}
\label{fig1}
\end{figure}

A natural question is whether an analogous boundary description of interior growth
can be developed beyond two-dimensional gravity. Already in AdS$_3$, however, a
qualitative difference arises: the codimension-one object whose growth enters the
CV proposal is now a surface rather than a geodesic, and its boundary consists of
spatial circles rather than isolated points. Consequently, its volume is not
naturally encoded in a two-point function of local boundary operators. In this work,
we propose a way to overcome this obstacle for the two-sided BTZ black hole. The Euclidean preparation of the thermofield double (TFD) state naturally lives on a thermal torus,
suggesting two extensions of the local AdS$_2$/JT construction. First, we replace
local primary operators by non-local operators obtained by smearing them along the
spatial $S^1$, reflecting the extended spatial support of the bulk surface. Second,
we further average the resulting correlator over appropriate segments of the
Euclidean thermal circle. This temporal averaging is compatible with KMS periodicity
and is analogous in spirit to the integrated correlators appearing in the
information-metric constructions of Refs.~\cite{Miyaji:2016fse,Miyaji:2015woj}.
The resulting two-point function in the TFD state provides a
higher-dimensional analog of the local correlator entering the AdS$_2$/JT
construction. With an appropriate spatial smearing prescription and Euclidean-time
integration, we find that its Lorentzian time dependence reproduces the
characteristic growth of the codimension-one surface extending through the BTZ
interior; see Fig.~\eqref{fig1}. Our observable is nevertheless distinct from those of
Refs.~\cite{Miyaji:2016fse,Miyaji:2015woj}: it involves a non-trivial spatial
smearing kernel adapted to the extended bulk surface and is not defined through a
Fisher-information-type overlap.

To evaluate this proposal, we also use the Chern--Simons formulation of pure
three-dimensional gravity, in which classical AdS$_3$ geometries are encoded in
flat gauge connections and their holonomies \cite{Castro:2016ehj}. Building on the
Wilson-line construction of Ref.~\cite{Castro:2016ehj}, we first evaluate the
fixed-endpoint Wilson line and take the asymptotic limit in which its endpoints lie
on the two boundaries of the eternal BTZ geometry. We then smear these endpoints
along the spatial circle and, in direct analogy, perform the same Euclidean-time
averaging introduced in the boundary construction. The resulting bulk calculation reproduces the expected late-time linear growth, providing a non-trivial check of our proposed reconstruction. This suggests an extension of
the familiar relation between bulk Wilson lines and holographic entanglement entropy
\cite{Ammon:2013hba} to codimension-one observables associated with complexity.

Importantly, our analysis is semiclassical and does not address the eventual non-perturbative saturation of the interior volume. This limitation is particularly relevant in three dimensions, where an analog of the matrix-model description underlying the non-perturbative completion of JT gravity is not presently available in comparable generality \cite{deBoer:2024mqg,DiUbaldo:2023qli,Boruch:2025ilr}. We therefore focus on the regime in which the semiclassical description is reliable, including the late-time linear-growth regime but parametrically before the exponentially long time scales at which non-perturbative effects become essential.

The second part of this work asks whether the same geometric growth is encoded in Krylov complexity. Krylov methods provide a useful framework for characterizing quantum dynamics by mapping time evolution onto an effective one-dimensional chain. See  \cite{Nandy:2024evd,Rabinovici:2025otw,Jeong:2026toappear} for reviews. For operator evolution, repeated action of the Liouvillian generates an orthonormal Krylov basis \cite{viswanath1994recursion,Parker:2018yvk,Muck:2022xfc}, with the associated Lanczos coefficients governing the spreading of the operator along the chain \cite{Parker:2018yvk}. The resulting operator Krylov complexity can be reconstructed from moments of a two-point function by solving the associated discrete Schr\"odinger equation \cite{Barbon:2019wsy}. An analogous construction applies to state evolution: repeated action of the Hamiltonian on an initial state defines a Krylov basis in Hilbert space, and the spreading of the evolved state over this basis gives the Krylov spread complexity \cite{Balasubramanian:2022tpr}. In this case, the relevant moments may be obtained from the analytic continuation of the Euclidean partition function \cite{Balasubramanian:2022tpr}. Both operator and spread complexity have found numerous applications in gravitational systems and quantum-chaotic many-body dynamics \cite{Dymarsky:2019elm, Dymarsky:2021bjq,Camargo:2022rnt,Bhattacharjee:2022vlt,ballar2022krylov,Bhattacharyya:2023dhp,Huh:2023jxt,Bhattacharya:2023zqt,Alishahiha:2022anw,Bhattacharjee:2022qjw,Erdmenger:2023wjg,Bhattacharya:2023yec,Nandy:2023brt,NSSrivatsa:2023pby,liu2023krylov,Beetar:2023mfn,Das:2024zuu,Caputa:2024vrn,Alishahiha:2024vbf,bento2024krylov,menzler2024krylov,Balasubramanian:2024ghv,Camargo:2024deu,Baggioli:2024wbz,Huh:2024ytz,Jeong:2024jjn,Murugan:2024ory,Aguilar-Gutierrez:2025hbf,Takahashi_2025,Bhattacharyya:2025gvd,Aguilar-Gutierrez:2026ogo,Chakrabarti:2025hsb,Baggioli:2025knt,Bhattacharyya:2025lsc,Balasubramanian:2026azk,Begines:2026fnx,Murugan:2026yyu,Bhattacharyya:2026qef,open1,Banerjee:2026kav,Murugan:2026rfa,FarajiAstaneh:2025thi,Grabarits:2026hjz,FarajiAstaneh:2026aks}.

These notions have acquired particularly concrete gravitational interpretations in two dimensions. In the triple-scaled limit of double-scaled SYK, the Krylov spread complexity of the infinite-temperature thermofield double state can be identified with the length of the two-sided wormhole using the chord-diagram formulation \cite{Rabinovici:2023yex}. This correspondence has subsequently been extended to finite temperature and to quantum corrections in double-scaled SYK and its sine-dilaton gravity dual \cite{Heller:2024ldz}. On the operator side, microcanonical Krylov complexity in the random-matrix description of JT gravity exhibits a post-scrambling regime of linear growth closely paralleling that of the maximal bulk volume \cite{Kar:2021nbm}. More recently, operator complexity of matter-excited thermofield double states in double-scaled SYK were shown to admit a direct geometric interpretation in terms of a bulk length operator \cite{Ambrosini:2024sre}. Beyond two-dimensional gravity, the holographic interpretation of Krylov complexity is less firmly established.\footnote{An important exception is provided by locally excited holographic CFT states, for which the growth rate of Krylov spread complexity can be related to the proper radial momentum of the corresponding infalling bulk excitation \cite{Caputa:2024sux}. This correspondence has since been explored and extended in a variety of higher-dimensional and non-conformal holographic settings \cite{Fan:2024iop,Aguilar-Gutierrez:2025kmw,Fu:2025kkh,Fatemiabhari:2025usn,Fatemiabhari:2026goj,Nunez:2026vhw}.}

Motivated by these developments, we investigate both notions of Krylov complexity in three-dimensional gravity. For operator Krylov complexity, we consider the same non-local operators that enter our reconstruction of the BTZ interior. From their correlation functions, we extract the spectral moments and corresponding Lanczos data within a microcanonical window. At large Krylov index, the Lanczos coefficients approach a constant, giving rise to linear growth of the operator Krylov complexity at late times. This behavior mirrors the characteristic linear growth of the codimension-one bulk surface. The coefficient of this growth may depend on the choice of smearing kernel, so the correspondence should not be interpreted as a universal equality of normalizations;\footnote{An intriguing possibility is that different choices of kernel may select different codimension-one observables within the broader complexity$=$anything framework \cite{Belin:2021bga,Belin:2022xmt,Myers:2024vve,Caceres:2025myu}, potentially providing a boundary realization of alternative geometric prescriptions. We leave this possibility for future investigation.} rather, the robust feature is the common linear time dependence. The smearing kernel is chosen to respect the periodicity of the spatial
circle and reduces to a local insertion in an appropriate limit.

The picture is markedly different for Krylov spread complexity. Starting
from the semiclassical gravitational partition function, we reconstruct the
Lanczos dynamics of the TFD state and find that, although the coefficients
become asymptotically $\mathfrak{su}(1,1)$-like, the resulting dynamics is
not described by an exact $\mathfrak{su}(1,1)$ chain. Instead, the spread
complexity remains approximately quadratic over the accessible time range,
with the numerical results strongly suggesting that this behavior persists
at late times. It therefore fails to reproduce the characteristic linear
growth of the bulk volume. This contrast constitutes one of our
main results: while operator Krylov complexity captures the characteristic
linear growth of the black hole interior, semiclassical Krylov spread
complexity does not. Our findings thus extend the connection between
operator growth and bulk geometry from two-dimensional models to
AdS$_3$, while highlighting an important distinction between operator and
state notions of Krylov complexity.

The remainder of the paper is organized as follows. In Sec.~(\ref{sec3}), we study the growth of the BTZ wormhole and introduce the smeared non-local operators used to reconstruct the codimension-one bulk surface, including their description in the Chern--Simons formulation of three-dimensional gravity. In Sec.~(\ref{secc5}), we develop the Krylov construction for these operators, determine their Lanczos data, and analyze the resulting operator complexity. Sec.~(\ref{secc4}) is devoted to the Krylov spread complexity of the thermofield double state, obtained from the semiclassical gravitational partition function in AdS$_3$. We conclude and discuss future directions in Sec.~(\ref{sec5}). Technical details concerning the smeared two-point functions, the relevant time integrals, the solution of the Krylov recursion relations, and TFD correlators are collected in appendices~(\ref{appAA}--\ref{sec:krylov_fluctuations}).

\section{Growth of the BTZ interior}\label{sec3}
In the probe approximation, the holographic two-point function of a
sufficiently heavy local operator, $1 \ll \Delta\ll c$, is related to the length of a
renormalized bulk geodesic connecting the corresponding boundary insertions
\cite{Balasubramanian:1999zv,Louko:2000tp}.
While such a relation is not specific to two dimensions, AdS$_2$/JT gravity has the
special feature that the codimension-one volume entering the Complexity$=$Volume
proposal reduces precisely to a length. For the two-sided black hole, this is the
length of the Einstein-Rosen bridge connecting the two boundaries. Consequently,
the same two-point function that probes the bulk geodesic also provides access to
the geometric quantity entering the CV proposal. Using the matrix-model description
of JT gravity, this relation can be studied non-perturbatively. In particular, the
non-perturbative contribution to the density-density correlator leads to saturation
of the geodesic length connecting the two boundaries \cite{Iliesiu:2021ari}.\footnote{For a recent extension to holographic setups in the presence of $T\bar{T}$ deformations, see
\cite{Bhattacharyya:2025gvd}.}

The goal of this section is to seek an analogous construction in three dimensions,
providing a boundary interpretation of the volume entering the CV proposal. In this
case, the relevant codimension-one object is a surface rather than a geodesic. As
illustrated in Fig.~\eqref{fig1}, for the two-sided BTZ black hole this surface is
anchored on spatial circles at the two asymptotic boundaries, unlike in AdS$_2$,
where the geodesic is anchored at isolated boundary points. This suggests that the
appropriate boundary observable should be spatially extended. At the same time, the
Euclidean preparation of the TFD state naturally involves the thermal circle,
motivating additional averaging over Euclidean time. Such integrated two-point
functions also arise in information-metric constructions
\cite{Miyaji:2016fse,Miyaji:2015woj}, providing further motivation for this
prescription. Motivated by these observations, and by the use of extended non-local
observables such as Wilson loops to probe bulk surfaces in holography
\cite{Maldacena:1998im,Berenstein:1998ij}, we consider non-local operators obtained
by smearing primary operators along the spatial $S^1$, and subsequently integrate
their two-point function over appropriate segments of the Euclidean thermal circle.

\begin{figure}[t!]
    \centering
\includegraphics[width=0.75\linewidth]{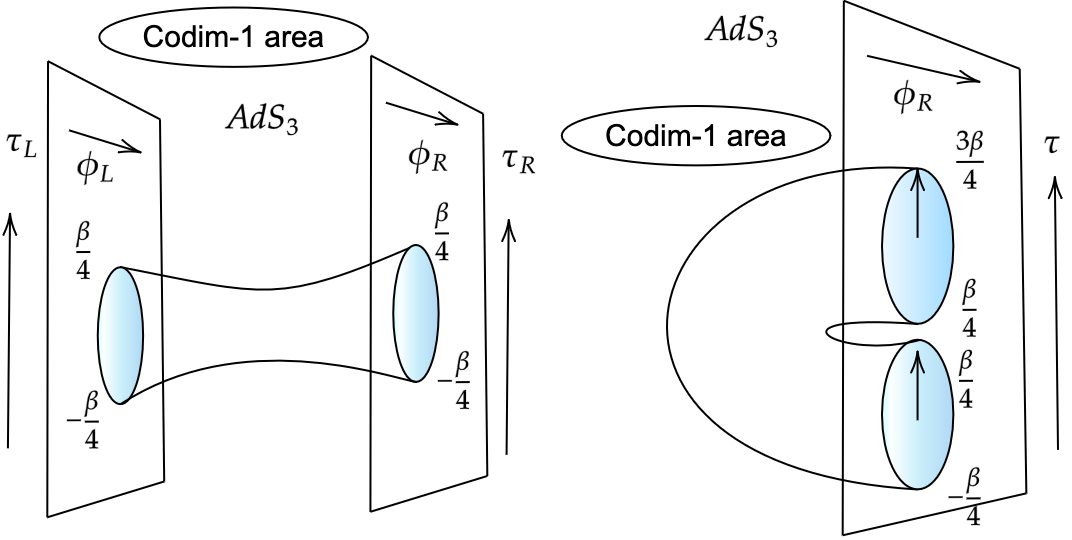}
    \caption{Integration regions for the Euclidean time coordinate $\tau$ associated with the extended codimension-one area in AdS$_3$. The left panel corresponds to the two-sided correlator, while the right panel shows the corresponding single-boundary correlator.}
    \label{fig2}
\end{figure}

The two-sided BTZ black hole is dual to a TFD state. We
therefore consider the two-point function of the smeared operators, with the two
insertions placed on opposite boundaries of the TFD state. Motivated by the role of
the two-point function as a probe of bulk geometry, we seek a
higher-dimensional relation of the form\footnote{The object on the left of \eqref{smearedopdef} should not be interpreted as a
correlator of boundary Wilson loops of the type considered in
\cite{Betzios:2023obs}. There, the non-local observables are genuine Wilson loops
in holographic gauge theories, whereas the two-dimensional CFT considered here is
not assumed to possess a microscopic gauge connection defining such an operator.
Our observable is instead defined directly in terms of spatially smeared primary
operators. The Wilson line introduced below should be understood as its bulk
Chern--Simons representation, rather than as a boundary Wilson-loop operator.}
\begin{equation}\label{smearedopdef}
    \int_{-\frac{\beta}{4}-\tau^E}^{\frac{\beta}{4}+\tau^E} d\tau_1^E\int_{-\frac{\beta}{4}-\tau^E}^{\frac{\beta}{4}+\tau^E} d\tau_2^E\langle \widetilde{\mathcal O}_{L}
    \widetilde{\mathcal O}_{R}\rangle_{\rm TFD}
    \sim\mathbfcal{A}^{L|R}(C)\,,
\end{equation}
where $\widetilde{\mathcal O}_{L,R}$ are spatially smeared operators and
$\mathbfcal{A}^{L|R}(C)$ denotes a codimension-one observable connecting the two
boundaries; see Fig.~\eqref{fig2}. The Euclidean-time integration in this construction deserves further explanation. Once the boundary insertion is spatially extended, it is no longer associated with a unique spacetime point and naturally acquires an extended domain
of dependence, motivating an additional averaging over Euclidean time. This
prescription is compatible with the KMS periodicity of the TFD state and is
analogous in spirit to the integrated correlators appearing in information-metric constructions \cite{Miyaji:2015woj,Miyaji:2016fse}, where overlaps between states
prepared by Euclidean path integrals lead to two-point functions integrated over
appropriate Euclidean-time regions. In the TFD geometry, these regions correspond
to segments of the thermal circle, with Lorentzian time dependence obtained by
analytic continuation.

In the following subsections, we formulate this boundary observable more explicitly
and evaluate it holographically using the Chern--Simons Wilson-line formalism of
Ref.~\cite{Castro:2016ehj}, incorporating both the spatial smearing and
Euclidean-time averaging. The resulting calculation reproduces the expected
late-time linear growth. This suggests an extension of the relation between
holographic entanglement entropy and bulk Wilson lines \cite{Ammon:2013hba} to
codimension-one observables associated with complexity.

\subsection{Correlation functions of non-local operators}\label{ssec21}
We now formulate the construction more explicitly. For the two-sided asymptotically
AdS$_3$ geometry, we consider a Euclidean-time-integrated two-point function of the
smeared operators discussed above,
    \begin{align}
    \begin{split}
    &\mathbfcal{A}^{L|R}(C)\xrightarrow{r\to \infty}\int_{-\frac{\beta}{4}-\tau^E}^{{\frac{\beta}{4}+\tau^E}}\int_{-\frac{\beta}{4}-\tau^E}^{{\frac{\beta}{4}+\tau^E}} d\tau^E_1d\tau^E_2\langle\Psi_{\text{TFD}}|\oint_{C} dx'K(x,x')\,\mathcal{O}_\Delta^{L}(x')\\&\hspace{8 cm}\times\oint_{C} dy'K(y,y')\,\mathcal{O}_\Delta^{R}(y')|\Psi_{\text{TFD}}\rangle \\&
    \hspace{1 cm}=\int_{-\frac{\beta}{4}-\tau^E}^{{\frac{\beta}{4}+\tau^E}}\int_{-\frac{\beta}{4}-\tau^E}^{{\frac{\beta}{4}+\tau^E}} d\tau^E_1d\tau^E_2\,\underbrace{\langle\Psi_{\text{TFD}}|\,,\begin{minipage}[h]{0.12\linewidth}
	\vspace{1pt}
	\scalebox{1}{\includegraphics[width=\linewidth]{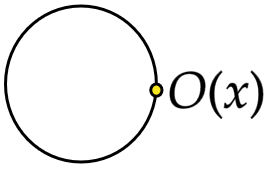}}
   \end{minipage}
   \begin{minipage}[h]{0.12\linewidth}
	\vspace{1pt}
	\scalebox{1}{\includegraphics[width=\linewidth]{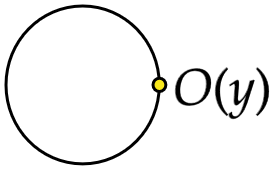}}
   \end{minipage}|\Psi_{\text{TFD}}\rangle}_{G_{\beta}^{\text{extended}}}\,.
    \label{3.2r}
    \end{split}
\end{align}
Here, $r\to\infty$ denotes the asymptotic boundary limit, with the two
operators inserted on opposite boundaries of the two-sided BTZ geometry. The
non-local operators are obtained by smearing a primary operator along the spatial
circle,
\begin{align}
    \begin{minipage}[h]{0.12\linewidth}
	\vspace{0pt}
	\scalebox{1}{\includegraphics[width=\linewidth]{OP.png}}
\end{minipage}\,=\oint dx'\underbrace{K(x,x')}_{\text{smearing kernel in the CFT}}\,\,\,\underbrace{\mathcal{O}_{\Delta}(x')}_{\text{Primary operator}}.\label{3.3r}
\end{align}
The expectation value in \eqref{3.2r} is evaluated in the TFD state
$|\Psi_{\rm TFD}\rangle$. The coordinates $x$ and $y$ denote spatial
boundary coordinates. The smearing in \eqref{3.3r} acts only along the
spatial circle, while the Euclidean-time averaging is implemented
separately through the integrations in \eqref{3.2r}. We denote the
circumference of the spatial circle by $L$; in the convention
$\phi\sim\phi+2\pi$ used for BTZ, $L=2\pi$.

The spatial smearing kernel is not uniquely fixed a priori. We choose a
periodic Gaussian kernel on the spatial circle,
\begin{align}
K(\phi,\phi')
=
\mathcal N_\sigma
\sum_{n\in\mathbb Z}
e^{-\sigma n^2}
e^{\frac{2\pi i n}{L}(\phi-\phi')} ,
\label{eq:smearing-kernel}
\end{align}
where $\sigma$ controls the smearing width and $\mathcal N_\sigma$ is an
overall normalization. This kernel is manifestly periodic under
$\phi\to\phi+L$ and $\phi'\to\phi'+L$. In the limit $\sigma\to0$, it
approaches the periodic delta function, up to the overall normalization,
and therefore recovers the corresponding local operator. Choosing
$\mathcal N_\sigma=1/L$ gives the conventionally normalized periodic
delta function. Since the overall normalization does not affect the time dependence of
the observables considered below, we suppress $\mathcal N_\sigma$ in what
follows.
    
By contrast, the Euclidean-time averaging is performed with a homogeneous
kernel. This prescription is consistent with the
Kubo--Martin--Schwinger (KMS) periodicity of the thermal state and with
the integrated-correlator construction motivated above. The spatial
smearing and Euclidean-time averaging are therefore independent
operations, controlled respectively by $K(\phi,\phi')$ and the
thermal-circle integrations.

\subsection{Chern--Simons representation}

To evaluate \eqref{3.2r} from the bulk, we make use of the Chern--Simons
formulation of pure three-dimensional gravity. This provides a natural framework
for relating the extended boundary correlator introduced above to a bulk Wilson
line whose endpoints are subsequently smeared over the boundary spatial circle.
Following \cite{Castro:2016ehj}, a Wilson line in an infinite-dimensional
representation $\mathbfcal{R}$, extending along a curve $C_{ij}$ between two bulk
points $y_i$ and $y_j$, can be written as
\begin{align}
    \mathbfcal{W}_{\mathbfcal{R}}(y_i,y_j)=\langle U_i|\mathscr{P}\,\exp\bigg(\int_{C_{ij}} A\bigg)\,\mathscr{P}\exp\bigg(\int_{C_{ij}} \bar A\bigg)|U_f\rangle\,.
\end{align}
Here, $U(y)$ denotes the auxiliary field living on the Wilson line, while
$A$ and $\bar A$ are the two Chern--Simons connections. When the endpoints are
pushed to the asymptotic boundary, the Wilson line reduces to the two-point
function of primary operators \cite{Castro:2016ehj},
\begin{align}
    \mathbfcal{W}_{\mathbfcal{R}}\xrightarrow[r\to \infty]{} \langle\Psi|O_{\Delta}(x_i)O_{\Delta}(x_j)|\Psi \rangle\,.
\end{align}
Here, $x_i$ and $x_j$ denote the boundary endpoints, while $|\Psi\rangle$ is
the heavy background state encoded by the bulk gauge connections. The operators
$\mathcal{O}_{\Delta}$ are light probes whose conformal dimension $\Delta$ is
kept fixed in the large-$c$ limit. In what follows, we specialize to the case in
which $|\Psi\rangle$ is the thermofield double state.

This construction directly applies to correlators of local boundary operators.
Our observable, however, involves the non-local operators introduced in \eqref{3.3r}. The corresponding bulk prescription is obtained by first
evaluating the Wilson line for fixed boundary endpoints and subsequently smearing
each endpoint along the spatial circle with the same kernel $K$.\footnote{For fixed
endpoints, the two-sided Wilson-line observable may be written schematically as
\begin{align}
\Tr_{\mathbfcal{R}}
\mathbfcal{W}_{\mathbfcal{R}}(x,y|x,y)
\sim
\langle\Sigma|\,
\Tr_{\mathbfcal{R}}
\underbrace{
\mathscr{P}e^{\int_{C_{xy}} a^L(x')}
}_{\Phi^L(x,y)}
\,
\Tr_{\mathbfcal{R}}
\underbrace{
\mathscr{P}e^{\int_{C_{xy}} \bar a^R(x')}
}_{\Phi^R(x,y)}
\,|\Sigma\rangle\, ,
\end{align}
where $|\Sigma\rangle$ is a reference state specifying the boundary conditions at
the endpoints, and $C_{xy}$ denotes a bulk path anchored at the boundary points
$x$ and $y$.}
Schematically, the smeared observable is therefore
\begin{align}
\widetilde{\mathbfcal W}^{L|R}(x,y)
=
\oint_C dx'\oint_C dy'\,
K(x,x')K(y,y')\,
\mathbfcal W^{L|R}(x',y')\, .
\label{2.10o}
\end{align}
This implements directly in the bulk the non-local boundary operators appearing
in \eqref{3.2r}. The spatial integration is the essential extension of the
fixed-endpoint Wilson-line construction of \cite{Castro:2016ehj}.

We now specialize to the non-rotating BTZ black hole. In the
$SL(2,\mathbb{R})\times SL(2,\mathbb{R})$ Chern--Simons formulation of
three-dimensional Einstein gravity, the metric can be written as
\cite{Banados:2015tft,Castro:2016ehj}
\begin{align}
ds^2
=
-e^{-2\rho}\left(e^{2\rho}-2\pi\mathcal L\right)^2dt^2
+e^{-2\rho}\left(e^{2\rho}+2\pi\mathcal L\right)^2d\phi^2
+d\rho^2\, .
\end{align}
The corresponding gauge connections take the form
\begin{align}
A
&=
b(r)^{-1}\bigl(a(x^+,x^-)+d\bigr)b(r),
\nonumber\\
\bar A
&=
\bar b(r)\bigl(\bar a(x^+,x^-)+d\bigr)\bar b(r)^{-1},
\label{3.7y}
\end{align}
where
\begin{align}
a(x^+,x^-)
&=
\left(L_1-2\pi\mathcal L L_{-1}\right)dx^+,
\nonumber\\
\bar a(x^+,x^-)
&=
-\left(L_{-1}-2\pi\mathcal L L_1\right)dx^-,
\qquad
x^\pm=t\pm\phi\, .
\end{align}
The radial coordinate is related to $\rho$ by
\begin{align}
r=e^\rho+2\pi\mathcal L\,e^{-\rho},
\end{align}
while the inverse temperature is given by
\begin{align}
\beta=\sqrt{\frac{\pi}{2\mathcal L}}\, .
\end{align}
The boundary connections $(a,\bar a)$ are flat and encode the boundary
dependence of the Chern--Simons fields. Since the bulk connections are flat, they can locally be expressed as pure
gauge. In the conventions
\begin{align}
A=\mathscr L\,d\mathscr L^{-1},
\qquad
\bar A=\mathscr R\,d\mathscr R^{-1},
\end{align}
a convenient choice for the left-moving group element is
\begin{align}
\mathscr L(y)
=
b(r)^{-1}
\exp\left(\int a_i\,dx^i\right),
\qquad
a=a_i dx^i\, .
\label{eq4.9}
\end{align}
The radial dependence is thus isolated in $b(r)$. Substituting
\eqref{eq4.9} into $\mathscr L\,d\mathscr L^{-1}$ reproduces the
connection in \eqref{3.7y} \footnote{Writing
$U\equiv\exp(\int a)$ and $\mathscr L=b^{-1}U$, one obtains
\begin{align}
\mathscr L\,d\mathscr L^{-1}
=
b^{-1}(U\,dU^{-1})b+b^{-1}db\, ,
\end{align}
which reproduces \eqref{3.7y} upon identifying
$U\,dU^{-1}=a$ in the conventions used here.}.
Similarly, for the right-moving sector we take
\begin{align}
\mathscr R(y)
=
\exp\left(-\int \bar a_i\,dx^i\right)\bar b(r)^{-1}\, .
\label{4.14r}
\end{align}
The gauge-invariant combination entering the Wilson-line observable is
\begin{align}
\mathbfcal M(y_i,y_f)
=
\mathscr R(y_i)\mathscr L(y_i)
\mathscr L(y_f)^{-1}\mathscr R(y_f)^{-1}\, ,
\label{2.19i}
\end{align}
where $y_i$ and $y_f$ denote the two bulk endpoints. We eventually take
$r\to\infty$, placing these endpoints on the asymptotic boundaries, and focus
on the configuration in which they lie on opposite sides of the two-sided
BTZ geometry.

We can now implement the spatial smearing discussed above. Evaluating the
fixed-endpoint Wilson line and subsequently integrating its two boundary
endpoints against the kernel $K$, the relevant codimension-one observable
takes the form
\begin{align}
\begin{split}
\mathbfcal A^{L|R}(C)
={}&
\int_{-\frac{\beta}{4}-\tau^E}^{\frac{\beta}{4}+\tau^E}
d\tau_1^E
\int_{-\frac{\beta}{4}-\tau^E}^{\frac{\beta}{4}+\tau^E}
d\tau_2^E
\oint_C dy_f^{\prime R}\,
K(y_f^{\prime R},y_f^R)
\oint_C dy_i^{\prime L}\,
K(y_i^{\prime L},y_i^L)
\\
&\,\,\,\,\,\,\times
\exp\left[
-\Delta\,
\cosh^{-1}\left(
\frac{1}{2}
\Tr\mathbfcal M(y_i^{\prime L},y_f^{\prime R})
\right)
\right] .
\label{4.17u}
\end{split}
\end{align}
Here we used the standard relation between the Chern--Simons Wilson line and
the group-element combination $\mathbfcal M$ \cite{Castro:2016ehj},
\begin{align}
\log\mathbfcal W_{\mathbfcal R}
=
-\Delta\,
\cosh^{-1}\left(
\frac{1}{2}\Tr\mathbfcal M
\right).
\end{align}
Importantly, the smearing is performed over the boundary endpoints after
evaluating the fixed-endpoint Wilson line. The kernels, therefore, multiply the
Wilson-line integrand and remain outside the $\cosh^{-1}$ in
\eqref{4.17u}.

\subsection{Computation of the left and right group elements}

Having established the smeared Wilson-line prescription, we now evaluate the group
elements entering \eqref{2.19i} explicitly for BTZ. The Euclidean regularity condition allows the temporal
components of the flat connections to be diagonalized as
\cite{Castro:2016ehj}
\begin{align}
a_\tau
=
Q\left(\frac{2\pi i}{\beta}L_0\right)Q^{-1},
\qquad
\bar a_\tau
=
\bar Q\left(\frac{2\pi i}{\beta}L_0\right)\bar Q^{-1},
\end{align}
so that, for a finite Euclidean-time interval,
\begin{align}
e^{a_\tau\Delta\tau_E}
=
Q
e^{\frac{2\pi i}{\beta}\Delta\tau_E L_0}
Q^{-1},
\end{align}
and analogously for the barred sector\footnote{Here we use the identity
$e^{QXQ^{-1}}=Qe^XQ^{-1}$.}.
The corresponding similarity matrices may be chosen as
\begin{align}
Q=
\left(
\begin{array}{cc}
-\frac{\pi}{\beta} & \frac{\pi}{\beta} \\
1 & 1
\end{array}
\right),
\qquad
\bar Q=
\left(
\begin{array}{cc}
-1 & 1 \\
\frac{\pi}{\beta} & \frac{\pi}{\beta}
\end{array}
\right),
\end{align}
up to an overall normalization. Here $L_0,L_{\pm1}$ denote the $\mathfrak{sl}(2,\mathbb R)$ generators,
corresponding to the global conformal subalgebra.

To evaluate the trace appearing in \eqref{4.17u}, it is convenient to
analytically continue to Lorentzian signature and work in Kruskal gauge.
The BTZ metric then takes the form
\begin{align}
ds^2
=
-\frac{4}{(1+uv)^2}du\,dv
+
\left(\frac{2\pi}{\beta}\right)^2
\left(\frac{uv-1}{uv+1}\right)^2d\phi^2 ,
\end{align}
where $u$ and $v$ are Kruskal light-cone coordinates and $\phi$ is the
spatial coordinate. The analytic continuation from the Euclidean
coordinates is
\begin{align}
w=-v\,,
\qquad
\bar w=u\,.
\end{align}
In this gauge, the spacetime-dependent group elements are
\cite{Castro:2016ehj}
\begin{align}
\begin{split}
\mathscr{L}(y)
&=
\left(
\begin{array}{cc}
\frac{\beta e^{-\frac{\pi\phi}{\beta}}
\left(e^{\frac{2\pi\phi}{\beta}}-w\right)}
{\sqrt{2\pi}\sqrt{\beta(1-w\bar w)}}
&
\frac{\sqrt{\frac{\pi}{2}}e^{-\frac{\pi\phi}{\beta}}
\left(e^{\frac{2\pi\phi}{\beta}}+w\right)}
{\sqrt{\beta(1-w\bar w)}}
\\[2mm]
\frac{\beta e^{-\frac{\pi\phi}{\beta}}
\left(\bar w e^{\frac{2\pi\phi}{\beta}}-1\right)}
{\sqrt{2\pi}\sqrt{\beta(1-w\bar w)}}
&
\frac{\sqrt{\frac{\pi}{2}}e^{-\frac{\pi\phi}{\beta}}
\left(\bar w e^{\frac{2\pi\phi}{\beta}}+1\right)}
{\sqrt{\beta(1-w\bar w)}}
\end{array}
\right),
\\[2mm]
\mathscr{R}(y)
&=
\left(
\begin{array}{cc}
\frac{\beta e^{-\frac{\pi\phi}{\beta}}
\left(e^{\frac{2\pi\phi}{\beta}}-\bar w\right)}
{\sqrt{2\pi}\sqrt{\beta(1-w\bar w)}}
&
\frac{\beta e^{-\frac{\pi\phi}{\beta}}
\left(w e^{\frac{2\pi\phi}{\beta}}-1\right)}
{\sqrt{2\pi}\sqrt{\beta(1-w\bar w)}}
\\[2mm]
\frac{\sqrt{\frac{\pi}{2}}e^{-\frac{\pi\phi}{\beta}}
\left(e^{\frac{2\pi\phi}{\beta}}+\bar w\right)}
{\sqrt{\beta(1-w\bar w)}}
&
\frac{\sqrt{\frac{\pi}{2}}e^{-\frac{\pi\phi}{\beta}}
\left(w e^{\frac{2\pi\phi}{\beta}}+1\right)}
{\sqrt{\beta(1-w\bar w)}}
\end{array}
\right).
\end{split}
\end{align}

To evaluate a Wilson line connecting the two asymptotic boundaries, we must specify
the group elements in the two exterior wedges of the maximally extended Lorentzian
geometry. In the right wedge, where $u>0$ and $v<0$, the Kruskal coordinates are
parametrized as
\begin{align}
u
=
\tanh\left(\frac{r}{2}\right)
e^{\frac{2\pi t_R}{\beta}},
\qquad
v
=
-\tanh\left(\frac{r}{2}\right)
e^{-\frac{2\pi t_R}{\beta}},
\label{2.28u}
\end{align}
whereas in the left wedge, where $u<0$ and $v>0$,
\begin{align}
u
=
-\tanh\left(\frac{r}{2}\right)
e^{-\frac{2\pi t_L}{\beta}},
\qquad
v
=
\tanh\left(\frac{r}{2}\right)
e^{\frac{2\pi t_L}{\beta}}.
\label{2.29o}
\end{align}
Here $t_L$ and $t_R$ denote the Lorentzian times on the left and right
boundaries, respectively. Using these parametrizations in \eqref{2.19i}, the group-theoretic invariant
entering the Wilson line becomes
\begin{align}
\begin{split}
\Tr\mathbfcal{M}(y_1,y_2)
={}&
\frac{e^{-\frac{2\pi(\phi_1+\phi_2)}{\beta}}}
{(1+u_1v_1)(1+u_2v_2)}
\Bigg[
4e^{\frac{2\pi(\phi_1+\phi_2)}{\beta}}
(u_1v_2+u_2v_1)
\\
&\qquad\qquad
+(u_1v_1-1)(u_2v_2-1)
\left(
e^{\frac{4\pi\phi_1}{\beta}}
+
e^{\frac{4\pi\phi_2}{\beta}}
\right)
\Bigg].
\end{split}
\end{align}
For $\phi_1=\phi_2$, this reduces to the standard fixed-endpoint BTZ
Wilson-line result of \cite{Castro:2016ehj}.

We next insert this result into the smeared Wilson-line prescription \eqref{4.17u} and take both endpoints to the asymptotic boundaries.\footnote{Because BTZ is obtained by a spatial identification, the
fixed-endpoint propagator contains a sum over spatial images, as reviewed in
Appendix~\eqref{appD}. In the high-temperature regime considered here, the leading
Rindler contribution gives the two-sided thermal correlator used in the main text.
We therefore retain only this leading contribution and neglect the remaining image
terms.} This gives
\begin{align}
\begin{split}
\mathbfcal{A}^{L|R}(C)
={}&
\int_{-\frac{\beta}{4}-\tau^E}^{\frac{\beta}{4}+\tau^E}
d\tau_1^E
\int_{-\frac{\beta}{4}-\tau^E}^{\frac{\beta}{4}+\tau^E}
d\tau_2^E
\oint_C
K(y_f^{\prime R},y_f^R)
\oint_C
K(y_i^{\prime L},y_i^L)
\\
&\times
\exp\Bigg\{
-\Delta
\lim_{\epsilon\to0}
\cosh^{-1}
\left[
\frac{N}{\epsilon^2}
\cosh\left(\frac{2\pi}{\beta}(t_L+t_R)\right)
+\xi
\right]
\Bigg\}.
\label{4.28u}
\end{split}
\end{align}
Here, $N$ is time independent, while
\begin{align}
\xi
=
\frac{N}{\epsilon^2}
\cosh\left[
\frac{2\pi}{\beta}(\phi_1-\phi_2)
\right].
\end{align}
Thus, the first term contains the entire Lorentzian time dependence and $\xi$
depends only on the spatial separation. In the asymptotic boundary limit $\epsilon\to0$, the argument of the inverse
hyperbolic cosine is large, so that $\cosh^{-1}x\simeq\log(2x)$. We therefore obtain
\begin{align}
\begin{split}
\mathbfcal{A}^{L|R}(C)
={}&
\int_{-\frac{\beta}{4}-\tau^E}^{\frac{\beta}{4}+\tau^E}
d\tau_1^E
\int_{-\frac{\beta}{4}-\tau^E}^{\frac{\beta}{4}+\tau^E}
d\tau_2^E
\oint_C
K(\phi_2^{\prime R},\phi_2^R)
\oint_C
K(\phi_1^{\prime L},\phi_1^L)
\\
&\times
\left[
\frac{N}{\epsilon^2}
\cosh\left(\frac{2\pi}{\beta}(t_L+t_R)\right)
+
\frac{N}{\epsilon^2}
\cosh\left(
\frac{2\pi}{\beta}
(\phi_1^{\prime L}-\phi_2^{\prime R})
\right)
\right]^{-\Delta}.
\label{2.36y}
\end{split}
\end{align}
Notice that the smearing kernels remain outside the $\cosh^{-1}$:
the fixed-endpoint Wilson line is evaluated first, and its boundary endpoints are subsequently integrated against the kernels. The resulting expression can be analyzed systematically at early and late times;
details of the integration are collected in Appendix~\eqref{appAA} (the spatial integrals) and Appendix~\eqref{appB} (the temporal integrals). For $\Delta=2$, the
late-time behavior is\footnote{Following Ref.~\cite{Carmi:2017jqz}, we introduce a length scale $R$ by regulating the spatial integral to $\phi\in[-2\pi R,2\pi R]$.
This makes the dependence on the spatial size explicit and allows us to track
the extensive divergence in the decompactification limit $R\to\infty$.}
\begin{align}
\mathbfcal{A}^{L|R}(C)
=
\frac{(2\pi R)\pi^2 t}{\beta^2\epsilon^4}
\left[
\coth\left(\frac{4\pi^2R}{\beta}\right)
-
\frac{4\pi^2R}{\beta}
\csch^2\left(\frac{4\pi^2R}{\beta}\right)
\right],
\qquad
t\gg\beta ,
\label{2.38k}
\end{align}
Thus, the smeared observable as illustrated in Fig.~(\ref{fig2}), grows linearly for $t\gg\beta\,.$ This is the characteristic late-time behavior of the
codimension-one surface entering the CV proposal. In the decompactification limit, $R\to\infty$ at fixed $\beta$, this
simplifies to
\begin{align}
\mathbfcal{A}^{L|R}(C)
=
\frac{(2\pi R)\pi^2}{\beta^2}\,t ,
\label{2.39i}
\end{align}
\begin{figure}
    \centering
\includegraphics[width=0.6\linewidth]{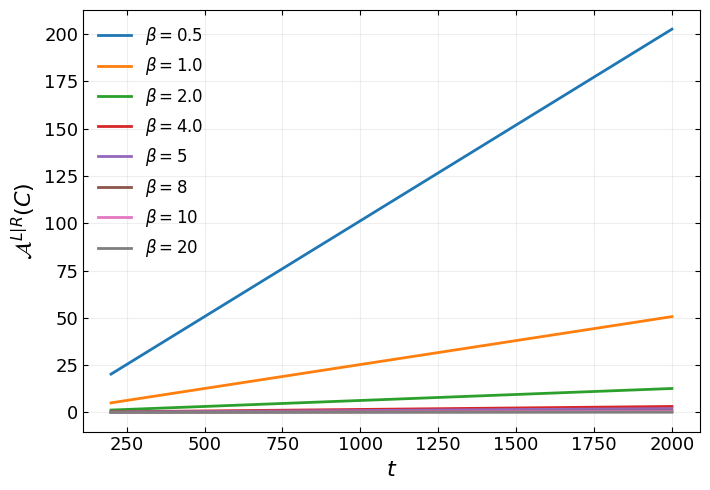}
    \caption{Late-time growth of the codimension-one observable
$\mathcal A^{L|R}(C)$, showing the asymptotically linear regime.}
    \label{fig:3}
\end{figure}
where we have suppressed the overall cutoff-dependent factor
$1/\epsilon^4$. The factor $2\pi R$ reflects the extensive divergence associated with the non-compact (after decompactifying) spatial direction. We plotted it in Fig.~\eqref{fig:3}. Thus,
$d\mathbfcal{A}^{L|R}/dt$ approaches a constant at late times, reproducing the characteristic linear time dependence of the codimension-one BTZ volume. The overall coefficient depends on the normalization and on the smearing prescription and should therefore not be interpreted as a universal equality with the CV growth rate.

Taken together, these results provide a three-dimensional extension of the
boundary-correlator picture familiar from two-dimensional gravity. The essential
new ingredient is the spatial smearing of the boundary insertions, which promotes
the fixed-endpoint Wilson line to an extended observable whose late-time behavior
reproduces the characteristic linear growth of the codimension-one BTZ surface.
From the bulk perspective, this suggests an extension of the familiar relation
between Wilson lines and codimension-two observables in holographic entanglement
entropy \cite{Ammon:2013hba} to extended observables associated with
complexity.

In the next section, we derive the spatially smeared two-point function
directly from the boundary CFT and use it to study the operator Krylov
complexity of the smeared operator. Working in a microcanonical window, we extract the corresponding Lanczos
data and show that the operator Krylov complexity develops a late-time linear
regime. We later contrast this behavior with the Krylov spread complexity of the
thermofield double state.

\section{Krylov operator complexity}\label{secc5}

We begin by briefly reviewing the Krylov construction for operator growth and the moment method used to determine the corresponding Lanczos coefficients. We also discuss how the Lanczos amplitudes can be obtained by solving the associated recursion relations. We then apply this framework to the extended operators introduced in Sec.~(\ref{sec3}) and compute their operator Krylov complexity, before contrasting the results with the Krylov spread complexity of the thermofield double state in Sec.~(\ref{secc4}).

\subsection{Preliminaries}

We begin by briefly reviewing the Krylov construction for operator evolution
\cite{Parker:2018yvk}. In the Heisenberg picture, the time
evolution of an operator $\mathcal{O}$ is governed by
\begin{align}
\partial_t \mathcal{O}(t)
=
i[H,\mathcal{O}(t)]\,,
\qquad
\mathcal{O}(t)
=
e^{iHt}\mathcal{O}(0)e^{-iHt}\,.
\end{align}
Introducing the Liouvillian superoperator,
\begin{align}
\mathcal{L} A \equiv [H,A]\,,
\end{align}
the Heisenberg evolution can equivalently be written as
\begin{align}
\begin{split}
\mathcal{O}(t)
&=
\mathcal{O}
+it[H,\mathcal{O}]
+\frac{(it)^2}{2!}[H,[H,\mathcal{O}]]
+\frac{(it)^3}{3!}[H,[H,[H,\mathcal{O}]]]
+\cdots
\\
&=
\sum_{n=0}^{\infty}
\frac{(it)^n}{n!}\mathcal{L}^n\mathcal{O}
=
e^{i\mathcal{L}t}\mathcal{O}\,.
\end{split}
\end{align}
The Krylov space associated with the initial operator $\mathcal{O}$ is
therefore the linear span of the operators generated by successive action of
the Liouvillian,
\begin{align}
\mathcal{H}_{\mathcal O}
=
\operatorname{span}
\left\{
\mathcal{O},
\mathcal{L}\mathcal{O},
\mathcal{L}^{2}\mathcal{O},
\ldots
\right\}.
\end{align}
Equivalently, it is the smallest subspace of an operator space that contains
$\mathcal{O}$ and is invariant under the action of $\mathcal{L}$. Since our application concerns thermal correlators in the TFD state, we equip
operator space with the thermal Wightman inner product,
\begin{align}
(A|B)_\beta
\equiv
\frac{1}{Z(\beta)}
\Tr\left(
e^{-\beta H/2}A^\dagger e^{-\beta H/2}B
\right).
\label{eq:thermal_inner_product}
\end{align}
In the doubled Hilbert-space representation, this inner product is equivalent to a two-sided TFD correlator, up to the conventional transpose of the operator acting on the left copy. This makes the connection with the correlators considered in Sec.~(\ref{sec3}) explicit.

Applying the Lanczos algorithm with respect to this inner product
orthonormalizes the sequence
$\{\mathcal O,\mathcal L\mathcal O,\mathcal L^2\mathcal O,\ldots\}$
and generates the Krylov basis $\{\mathcal O_n\}$. In general, the action of
the Liouvillian takes the tridiagonal form
\begin{align}
\mathcal L\mathcal O_n
=
b_{n+1}\mathcal O_{n+1}
+
a_n\mathcal O_n
+
b_n\mathcal O_{n-1}\,,
\qquad
b_0=0\,.
\end{align}
For a Hermitian seed operator, the thermal spectral measure associated with
the inner product \eqref{eq:thermal_inner_product} is symmetric under
$\omega\to-\omega$. Consequently, all odd moments vanish and the diagonal
Lanczos coefficients satisfy $a_n=0$. The recursion therefore reduces to
\begin{align}
\mathcal L\mathcal O_n
=
b_{n+1}\mathcal O_{n+1}
+
b_n\mathcal O_{n-1}\,,
\qquad
b_0=0\,,
\end{align}
where the positive coefficients $b_n$ are the Lanczos coefficients. The time-evolved operator can be expanded in the Krylov basis as
\begin{align}
\mathcal{O}(t)
=
\sum_{n=0}^{\infty}
i^n\phi_n(t)\mathcal{O}_n\,,
\end{align}
where $\phi_n(t)$ are the Krylov amplitudes. For a normalized initial
operator,
\begin{align}
\phi_n(0)=\delta_{n0}\,,
\qquad
\sum_{n=0}^{\infty}|\phi_n(t)|^2=1\,.
\end{align}
Substituting the Krylov expansion into the Heisenberg equation gives the
nearest-neighbor recursion relation
\begin{align}
\dot{\phi}_n(t)
=
b_n\phi_{n-1}(t)
-
b_{n+1}\phi_{n+1}(t)\,,
\qquad
\phi_{-1}(t)=0\,.
\label{eq:krylov_recursion}
\end{align}
The dynamics can therefore be interpreted as a hopping problem on a
semi-infinite one-dimensional chain, with hopping amplitudes set by
the Lanczos coefficients $b_n$.

Operator Krylov complexity measures the mean position of the evolving
wavefunction along this chain and is defined as
\begin{align}
\mathcal{K}_{\mathcal{O}}(t)
=
\sum_{n=0}^{\infty}
n\,|\phi_n(t)|^2\,.
\label{eq:operator_KC}
\end{align}
It therefore measures the mean distance traveled by the evolving operator
along the Krylov chain. For our purposes, it is particularly useful that the Lanczos coefficients
can be reconstructed directly from correlation functions. Defining the
thermal autocorrelation function
\begin{align}
C_\beta(t)
&\equiv
\bigl(\mathcal O_0\big|\mathcal O(t)\bigr)_\beta
=
\bigl(\mathcal O_0\big|
e^{i\mathcal Lt}
\big|\mathcal O_0\bigr)_\beta ,
\end{align}
its short-time expansion takes the form
\begin{align}
C_\beta(t)
=
\sum_{n=0}^{\infty}
\frac{(it)^n}{n!}\,\mu_n,
\qquad
\mu_n
=
\bigl(\mathcal O_0\big|
\mathcal L^n
\big|\mathcal O_0\bigr)_\beta .
\end{align}
Equivalently, the moments are moments of the corresponding spectral measure,
\begin{align}
\mu_n=\int d\omega\,\rho_{\mathcal O}(\omega)\,\omega^n .
\end{align}
For the symmetric spectral measure relevant here, all odd moments vanish,
and the sequence of even moments determines the Lanczos coefficients through
the standard moment--Lanczos construction
\cite{Parker:2018yvk,Barbon:2019wsy}.

\begin{figure} 
\centering \includegraphics[width=0.48\linewidth]{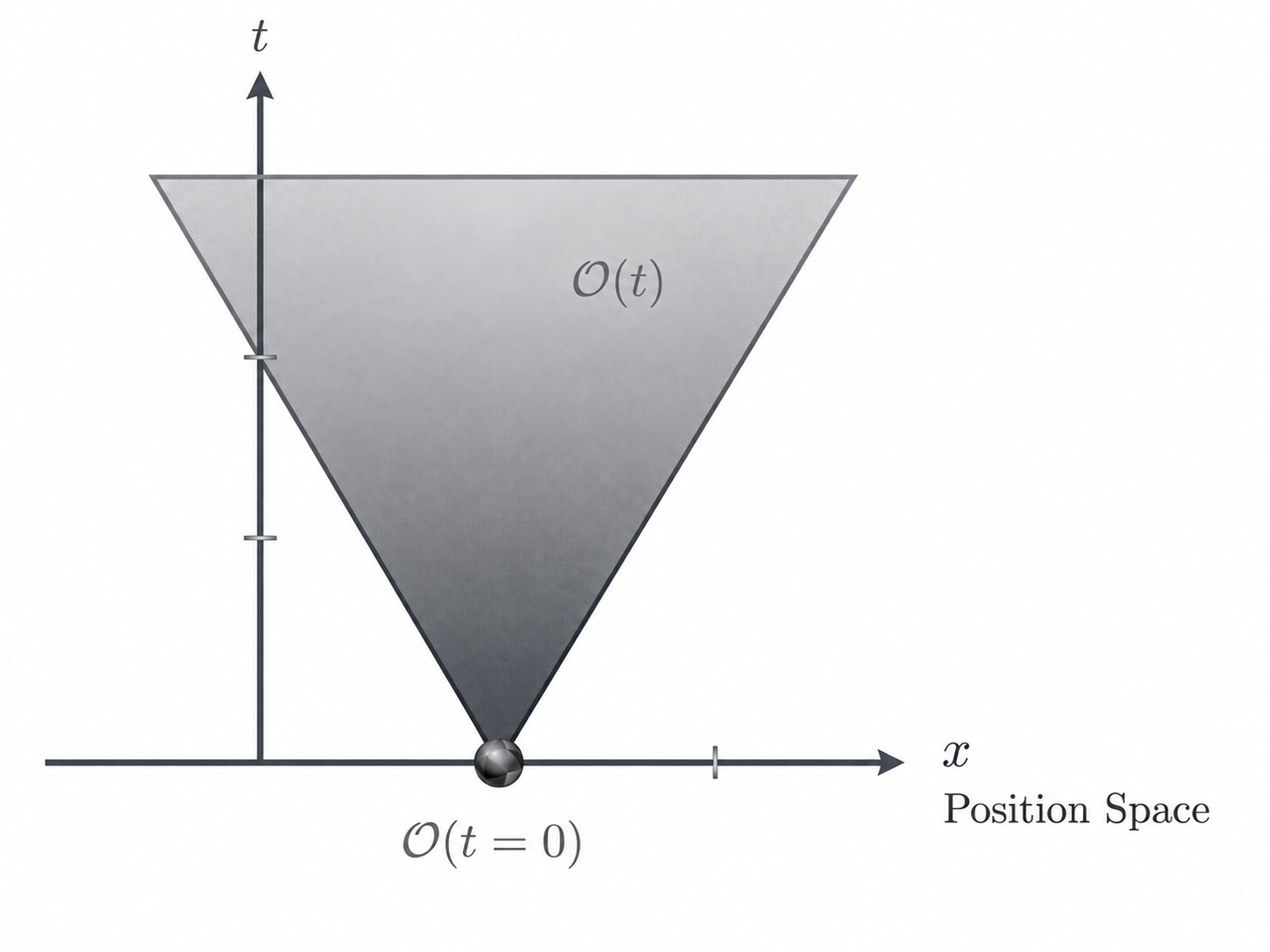} \includegraphics[width=0.48\linewidth]{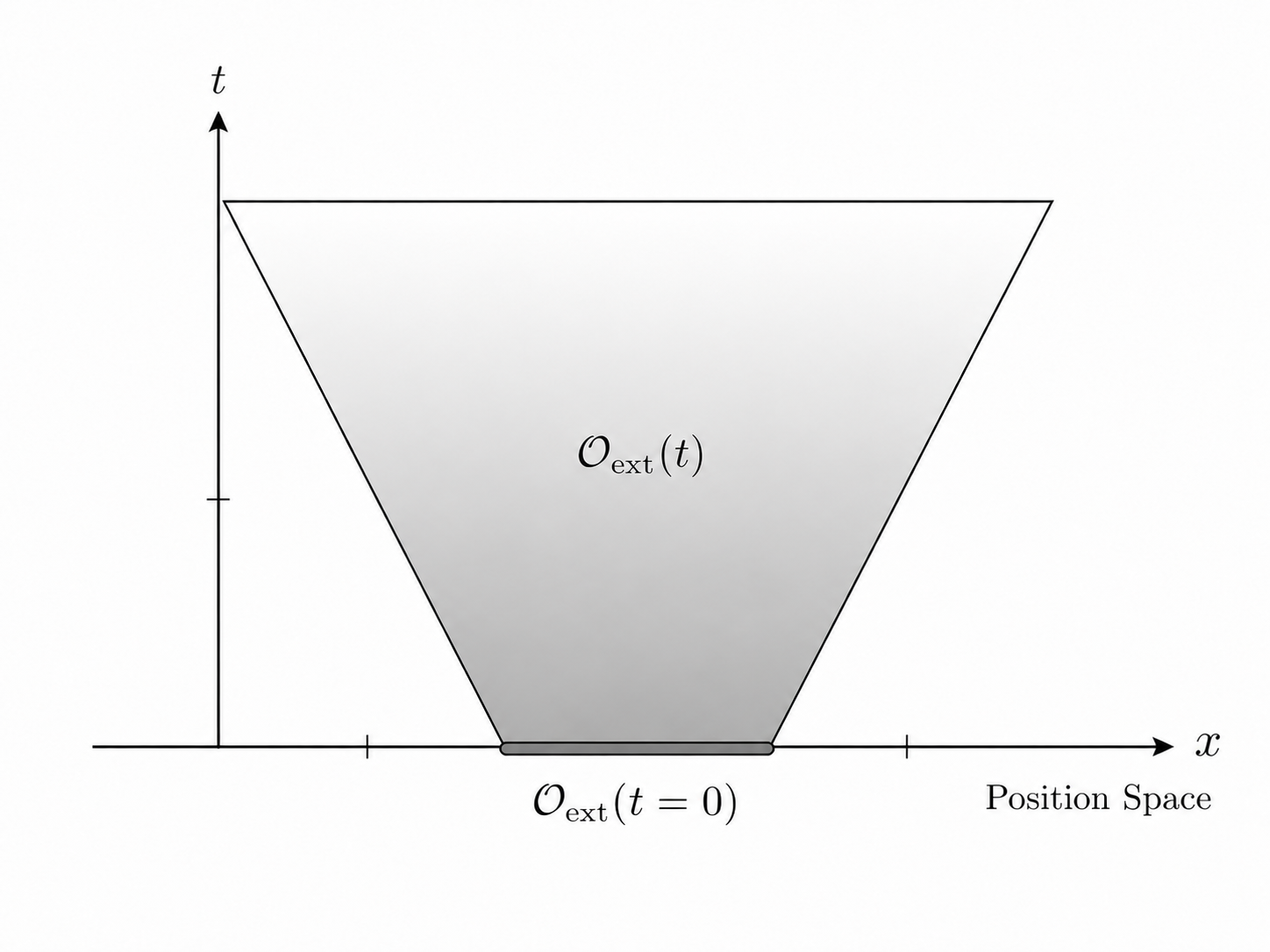} \caption{Schematic illustration of operator spreading for an initially
local operator (left) and an extended non-local operator (right). A local
operator is initially spatially localized, whereas the extended operator
already has nontrivial spatial support at $t=0$. This distinction concerns
the spatial structure of the seed operator in configuration space; in the Krylov representation, either operator defines a single initial basis element $\mathcal{O}_0$, and their subsequent dynamics is encoded in the corresponding Lanczos
data.} \label{fig:operator_growth} \end{figure}

The Krylov construction reviewed above applies equally to local and non-local
initial operators. The distinction lies in the spatial structure of the seed
operators: while a local operator is initially spatially localized, the
extended operators considered below already have nontrivial spatial support,
as illustrated schematically in Fig.~(\ref{fig:operator_growth}). In either
case, however, the chosen initial operator defines the first Krylov basis
element $\mathcal O_0$ of the chain, and the subsequent construction proceeds
in the same way.

This provides the strategy that we employ below. We first determine the thermal autocorrelation function of the spatially smeared operators introduced in Sec.~(\ref{sec3}). From it, we extract the corresponding fixed-energy spectral measure and its moments, which determine the asymptotic Lanczos coefficients and, ultimately, the operator Krylov
dynamics.

\subsection{Computation of operator complexity}
\label{3.2rr}

We now apply the Krylov construction to the extended operators
introduced in Sec.~(\ref{sec3}). For fixed Lorentzian times, we define the extended operators by smearing
the local primary along the spatial circle,\footnote{Extending the Krylov construction to include the Euclidean-time
integration of \eqref{3.2r} requires additional care, since a correlator
with time-dependent integration limits cannot automatically be interpreted
as the autocorrelation function of a fixed seed operator. One possibility
is to define the seed by convolution with a fixed filter $f(\tau)$ whose
integration domain is independent of the Lorentzian evolution time and,
for the microcanonical inverse-Laplace construction used below, of $\beta$.
The resulting spectral measure would then be weighted by the nonnegative
modulus squared of the corresponding transform of $f$. We leave
a systematic exploration of this possibility for future work.}
\begin{align}
\widetilde{\mathcal O}_{\Delta}^{L}(\phi_1,t_L)
&\equiv
\oint_C d\phi_1'\,
K(\phi_1,\phi_1')\,
\mathcal O_{\Delta}^{L}(\phi_1',t_L),
\nonumber\\
\widetilde{\mathcal O}_{\Delta}^{R}(\phi_2,t_R)
&\equiv
\oint_C d\phi_2'\,
K(\phi_2,\phi_2')\,
\mathcal O_{\Delta}^{R}(\phi_2',t_R).
\end{align}
Their thermal autocorrelation function in the TFD state is therefore
\begin{align}
\begin{split}
G_{\beta}^{\rm extended}
(t;\phi_1,\phi_2)
&\equiv
\langle\Psi_{\rm TFD}|
\widetilde{\mathcal O}_{\Delta}^{L}(\phi_1,t_L)
\widetilde{\mathcal O}_{\Delta}^{R}(\phi_2,t_R)
|\Psi_{\rm TFD}\rangle
\\
&=
\oint_C d\phi_1'\,
K(\phi_1,\phi_1')
\oint_C d\phi_2'\,
K(\phi_2,\phi_2')
\\
&\qquad\times
\langle\Psi_{\rm TFD}|
\mathcal O_{\Delta}^{L}(\phi_1',t_L)
\mathcal O_{\Delta}^{R}(\phi_2',t_R)
|\Psi_{\rm TFD}\rangle ,
\label{4.7i}
\end{split}
\end{align}
with $t\equiv t_L+t_R$.  The local two-sided correlator entering the last line of~\eqref{4.7i} is reviewed in Appendix~\eqref{appD}. In the high-temperature regime, analytic continuation
from Euclidean signature gives
\begin{align}
\langle
\mathcal O_{\Delta}^{L}(\phi_1',t_L)
\mathcal O_{\Delta}^{R}(\phi_2',t_R)
\rangle_{\beta}
=
\left(\frac{\pi}{\beta}\right)^{2\Delta}
\left[
\frac{N}{\epsilon^2}
\left\{
\cosh\left(\frac{2\pi t}{\beta}\right)
+
\cosh\left[
\frac{2\pi}{\beta}(\phi_1'-\phi_2')
\right]
\right\}
\right]^{-\Delta}.
\label{eq38app}
\end{align}
Plugging this into \eqref{4.7i} and using the kernel introduced in Sec.~(\ref{sec3}) we obtain
\begin{align}
\begin{split}
G_{\beta}^{\rm extended}
(\phi_1^L,\phi_2^R)
&=
\oint_C d\phi_2^{\prime R}\,
K(\phi_2^{\prime R},\phi_2^R)
\oint_C d\phi_1^{\prime L}\,
K(\phi_1^{\prime L},\phi_1^L)
\\
&\quad\times
\left[
\frac{N}{\epsilon^2}
\left\{
\cosh\left[\frac{2\pi}{\beta}(t_L+t_R)\right]
+
\cosh\left[
\frac{2\pi}{\beta}
(\phi_1^{\prime L}-\phi_2^{\prime R})
\right]
\right\}
\right]^{-\Delta}.
\end{split}
\end{align}
Expanding the smearing kernels in spatial Fourier modes and performing the
spatial integrations gives (see Appendix~\eqref{appAA} for details)
\begin{align}
\begin{split}
G_{\beta}^{\rm extended}
&=
\sum_{n=-\infty}^{\infty}
e^{\frac{2\pi i n}{L}(\phi_2-\phi_1)}
e^{-2\sigma n^2}
\frac{2^{2\Delta-1}\beta}{\pi}\Bigg[
\frac{u_+^{\lambda_n}}{\lambda_n}
F_1\left(
\lambda_n;
\Delta,\Delta;
\lambda_n+1;
-u_+e^{-2\pi t/\beta},
-u_+e^{2\pi t/\beta}
\right)
\\
&\hspace{3cm}
-
\frac{u_-^{\lambda_n}}{\lambda_n}
F_1\left(
\lambda_n;
\Delta,\Delta;
\lambda_n+1;
-u_-e^{-2\pi t/\beta},
-u_-e^{2\pi t/\beta}
\right)
\Bigg],
\label{3.10p}
\end{split}
\end{align}
where $F_1(a,b_1,b_2;c;x,y)$ is the Appell function defined by
\begin{align}
F_1(a,b_1,b_2;c;x,y)
=
\sum_{m=0}^{\infty}
\sum_{k=0}^{\infty}
\frac{(a)_{m+k}(b_1)_m(b_2)_k}
     {(c)_{m+k}\,m!\,k!}
x^m y^k ,
\end{align} 
and, following the conventions of Appendix~\eqref{appAA}, we have introduced
\begin{align}
\lambda_n=\Delta+\frac{i\beta n}{L},
\qquad
u_\pm=
\exp\left(\pm\frac{4\pi^2R}{\beta}\right).
\end{align}

\paragraph{Microcanonical spectral measure.}

To extract the Lanczos data, it is convenient to organize the correlator in
terms of the average energy and energy difference,
\begin{align}
E=\frac{E_i+E_j}{2},
\qquad
\omega=E_i-E_j,
\end{align}
following the microcanonical construction of \cite{Kar:2021nbm}. For a two-sided TFD correlator, the corresponding
spectral decomposition has the schematic form
\begin{align}
\widehat G_{\beta}(t)
\equiv
Z(\beta)G_{\beta}(t)
=
\sum_{i,j}
e^{-\frac{\beta}{2}(E_i+E_j)}
|\langle i|\mathcal O|j\rangle|^2
e^{i(E_i-E_j)t},
\end{align}
where we have absorbed the cylinder Casimir-energy shift into the definition
of the energies. Introducing the operator-weighted spectral measure
\begin{align}
\mathcal U(E,\omega)
=
\sum_{i,j}
|\langle i|\mathcal O|j\rangle|^2
\delta\left(
E-\frac{E_i+E_j}{2}
\right)
\delta\left(
\omega-(E_i-E_j)
\right),
\label{eq:spectral_measure}
\end{align}
we may equivalently write
\begin{align}
\widehat G_{\beta}(t)
=
\int_0^\infty dE\,e^{-\beta E}
\int_{-2E}^{2E}d\omega\,
\mathcal U(E,\omega)e^{i\omega t}.
\label{eq:spectral_decomposition}
\end{align}
The range $|\omega|\leq2E$ follows from the lower bound on the energy
spectrum.

The same construction applies to the extended operator. 
The corresponding operator-weighted spectral measure is then obtained by an inverse
Laplace transform in $\beta$ and an inverse Fourier transform in $t$,
\begin{align}
\begin{split}
\mathcal U^{\rm extended}(E,\omega)
&=
\mathcal L_{\beta\rightarrow E}^{-1}
\mathcal F_{t\rightarrow\omega}^{-1}
\left[
Z(\beta)\,
G_{\beta}^{\rm extended}
\right]
\\
&=
\int_{\gamma-i\infty}^{\gamma+i\infty}
\frac{d\beta}{2\pi i}\,
e^{\beta E}
\int_{-\infty}^{\infty}
\frac{dt}{2\pi}\,
e^{-i\omega t}
\,\cdots ,
\label{4.14r}
\end{split}
\end{align}
where the ellipsis denotes the expression obtained from \eqref{3.10p}. The explicit form is cumbersome for generic
$\Delta$, but simplifies considerably for particular values. For the analytic evaluation below, we specialize to $\Delta=1$.\footnote{In Sec.~\eqref{ssec21}, we analyzed the late-time behavior of the
non-local two-point function for $\Delta=2$. For the spectral moments, we instead
set $\Delta=1$, which allows greater analytic control over the resulting integral.
We have verified that this choice does not affect the qualitative late-time behavior
of the operator complexity. For $\Delta=2$, the corresponding integral is more
cumbersome, but the large-order moments exhibit the same edge-dominated scaling,
leading again to asymptotically constant Lanczos coefficients and linear late-time
operator complexity.} If we further restrict to the zero spatial Fourier mode, $n=0$, the Appell
function reduces to
\begin{align}
F_1
\left(
1;1,1;2;x,y
\right)
=
-\frac{\log\left(\frac{x-1}{y-1}\right)}{x-y}.
\end{align}
In this sector, Eq. \eqref{4.14r} reduces to,
\begin{align}
\begin{split}
\mathcal I(E,\omega)
&=
\int_{\gamma-i\infty}^{\gamma+i\infty}
\frac{d\beta}{2\pi i}\,
e^{\beta E}
\int_{-\infty}^{\infty}
\frac{dt}{2\pi}\,
e^{-i\omega t} \frac{1}{2\sinh(2\pi t/\beta)}
\\
&\quad\times
\left[
\log\left(
\frac{u_-e^{-2\pi t/\beta}+1}
     {u_-e^{2\pi t/\beta}+1}
\right)
-
\log\left(
\frac{u_+e^{-2\pi t/\beta}+1}
     {u_+e^{2\pi t/\beta}+1}
\right)
\right] .
\label{3.16y}
\end{split}
\end{align}
Evaluating the Fourier and inverse Laplace transforms gives, in the
small-frequency regime,
\begin{align}
\begin{split}
\mathcal{I}(E,\omega)
&=\frac{1}{8}
\int_{\gamma-i\infty}^{\gamma+i\infty}
\frac{d\beta}{2\pi i}\,
\frac{e^{\beta E}}{\beta}\,
\operatorname{sech}^{2}\!\left(
\frac{\beta|\omega|}{4}
\right),\\&
\approx
\frac{E}{2|\omega|}\,,
\qquad
0<|\omega|\ll E .
\label{3.17u}
\end{split}
\end{align}
This last expression should be understood as an intermediate-frequency
approximation, with the full spectral measure providing the appropriate
infrared regularization as $\omega\to0$. Moreover, the function multiplying $e^{-i\omega t}$ in the Fourier
transform \eqref{3.16y} is even under $t\to-t$, implying
\begin{equation}
    \mathcal I(E,\omega)=\mathcal I(E,-\omega)\,.
\end{equation}
Consequently, all odd moments vanish and the diagonal Lanczos coefficients
$a_n$ can be set to zero. We therefore focus below on the nonvanishing even
moments.

\paragraph{Moments and operator Krylov complexity.}
We can now determine the large-$n$ behavior of the Lanczos coefficients
from the moments of the microcanonical spectral measure. At fixed average
energy $E$, the normalized even moments are
\begin{align}
\mu_{2n}^{(E)}
=
\frac{1}{\mathcal N(E)}
\int_{-2E}^{2E}
d\omega\,
\omega^{2n}\,
\mathcal I(E,\omega),
\qquad
\mathcal N(E)
=
\int_{-2E}^{2E}
d\omega\,
\mathcal I(E,\omega),
\label{eq:moments_extended}
\end{align}
so that $\mu_0^{(E)}=1$. In the large-$n$ limit, assuming that the spectral measure is sufficiently
regular and has support extending to the endpoints $|\omega|=2E$, the
moments are dominated by the edges of the microcanonical interval.
Consequently,
\begin{align}
\mu_{2n}^{(E)}
\sim
A_n\,(2E)^{2n},
\qquad
n\gg1,
\label{eq:moment_asymptotics}
\end{align}
where $A_n$ is subexponential in $n$ and, within this approximation,
is given by
\begin{align}
   A_n=\frac{ 2 }{2n+1}(1-2^{1-2n})\,\zeta(2n)\,.
\end{align}
Notice that $A_n$ is dimensionless and independent of $E$. It arises
from the contributions near the edges of the microcanonical window,
$\omega\to\pm 2E$, obtained by carefully treating the full integral
in \eqref{3.17u}.

The edge-dominated growth in \eqref{eq:moment_asymptotics} is consistent with a Lanczos
plateau. More precisely, for a sufficiently regular spectral measure with
support on $[-2E,2E]$, the asymptotic recurrence coefficients approach
\begin{equation}
    b_n^{(E)}\longrightarrow b_\infty^{(E)}=E\,,
    \qquad n\to\infty\,.
\label{eq:lanczos_plateau_ext}
\end{equation}
A constant Lanczos plateau in turn leads to ballistic propagation along the Krylov chain and hence to linear growth of the operator Krylov complexity,
\begin{align}
\mathcal K_{\mathcal O}^{(E)}(t)
\propto
b_\infty^{(E)}\,t,
\qquad
t\gg b_\infty^{-1}.
\label{eq:linear_operator_KC}
\end{align}
The robust feature relevant for our purposes is the linear late-time
dependence, which mirrors the corresponding growth of the codimension-one
bulk observable. In the zero-mode sector analyzed explicitly above, the
dependence on the smearing width drops out. Dependence on the detailed
smearing profile can arise once the nonzero spatial Fourier modes are
retained, and we leave a systematic analysis of this dependence for future
work.

\paragraph{Comparison with local operators in JT gravity.}

It is instructive to contrast this result with the microcanonical operator
Krylov complexity studied in JT gravity \cite{Kar:2021nbm}. There, the
operator-weighted spectral measure of a local probe leads first to a
Lanczos ascent, associated with the scrambling regime, followed by a long
Lanczos plateau once the finite edges of the microcanonical spectrum become
important. The latter gives rise to the characteristic linear regime of
operator Krylov complexity.

The large-$n$ regime found here is somewhat different. For the extended
operators considered in AdS$_3$, the moments in the regime analyzed above
are already dominated by the endpoints of the microcanonical interval.
This directly leads to an asymptotically constant Lanczos sequence and,
consequently, linear Krylov growth. Thus, while a linear regime occurs in
both cases, the way in which the operator-weighted spectral measure
approaches the Lanczos plateau is different. It would be interesting to
understand to what extent this difference can be attributed specifically
to the non-local nature of the operators considered here; a more direct
analysis of their matrix elements would be required to establish this
connection.

In the next section, we turn to the Krylov spread complexity of the TFD
state. This is conceptually distinct from the operator construction above:
operator Krylov complexity describes Heisenberg evolution generated by the
Liouvillian, whereas spread complexity characterizes the evolution of a
state under repeated action of the Hamiltonian. We study the latter to
determine whether this state-based notion of Krylov complexity also captures
the characteristic growth of the two-sided BTZ interior.

\section{Krylov spread complexity}
\label{secc4}
In this section, we turn to the computation of Krylov spread complexity. In contrast to the operator Krylov complexity, spread complexity characterizes the evolution of a state under
repeated action of the Hamiltonian. We apply this construction to the TFD
state describing the two-sided BTZ black hole and investigate whether the
resulting state complexity captures the characteristic growth of the
black hole interior.

\subsection{Preliminaries}\label{sub41}

Let us first review the Krylov construction for state evolution
\cite{Balasubramanian:2022tpr}. Given a normalized initial state
$|\psi(0)\rangle$, repeated action of the Hamiltonian generates the Krylov
subspace
\begin{align}
\mathcal K
=
\operatorname{span}
\left\{
|\psi(0)\rangle,
H|\psi(0)\rangle,
H^2|\psi(0)\rangle,
\ldots
\right\}.
\label{eq:state_krylov_space}
\end{align}
The Lanczos algorithm then produces an orthonormal
basis $\{|K_n\rangle\}$, with
\begin{align}
|K_0\rangle=|\psi(0)\rangle\,,
\qquad
\langle K_m|K_n\rangle=\delta_{mn}\,.
\end{align}

The Lanczos construction can be implemented recursively by defining
\begin{align}
|A_{n+1}\rangle
=
(H-a_n)|K_n\rangle
-
b_n|K_{n-1}\rangle\,,
\label{eq:LanczosRecursionElab}
\end{align}
with
\begin{align}
a_n=
\langle K_n|H|K_n\rangle\,,\qquad
b_{n+1}=
\sqrt{\langle A_{n+1}|A_{n+1}\rangle}\,,
\qquad
|K_{n+1}\rangle
=
\frac{1}{b_{n+1}}|A_{n+1}\rangle\,,
\label{4.7y}
\end{align}
and initial conditions
\begin{align}
b_0=0\,,
\qquad
|K_{-1}\rangle=0\,.
\end{align}
The coefficients $a_n$ and $b_n$ are the Lanczos coefficients associated
with the pair $(H,|\psi(0)\rangle)$. In this basis, the Hamiltonian takes
the tridiagonal form
\begin{align}
H|K_n\rangle
=
a_n|K_n\rangle
+
b_{n+1}|K_{n+1}\rangle
+
b_n|K_{n-1}\rangle\,.
\label{eq:H_tridiagonal_elab}
\end{align}
Thus, as in the operator construction of Sec.~(\ref{secc5}), the original dynamics is
mapped onto an effective semi-infinite one-dimensional chain. The
coefficients $b_n$ play the role of nearest-neighbor hopping amplitudes,
while $a_n$ act as on-site potentials.

The time-evolved state can be expanded in the Krylov basis as
\begin{align}
|\psi(t)\rangle
=
\sum_{n=0}^{\infty}
\psi_n(t)|K_n\rangle\,.
\label{eq:state_krylov_expansion}
\end{align}
For a normalized initial state,
\begin{align}
\psi_n(0)=\delta_{n0}\,,
\qquad
\sum_{n=0}^{\infty}|\psi_n(t)|^2=1\,.
\end{align}
Substituting \eqref{eq:state_krylov_expansion} into the Schr\"odinger
equation gives
\begin{align}
i\partial_t\psi_n(t)
=
a_n\psi_n(t)
+
b_{n+1}\psi_{n+1}(t)
+
b_n\psi_{n-1}(t)\,,
\label{5.3r}
\end{align}
with $\psi_{-1}(t)=0$. The Krylov spread complexity is then defined as the mean position of the
state along the Krylov chain,
\begin{align}
\mathcal K_{\rm S}(t)
=
\sum_{n=0}^{\infty}
n\,|\psi_n(t)|^2 .
\label{eq:spread_complexity}
\end{align}

\paragraph{Survival amplitude and moments.}

The Lanczos data can be reconstructed directly from the moments of the
spectral measure associated with the initial state
\cite{Balasubramanian:2022tpr}. These moments are conveniently
obtained from the survival amplitude. For the TFD state, which we take as
the seed state in what follows, we define
\begin{align}
S_\beta(t)
\equiv
\langle\psi_\beta(t)|\psi_\beta(0)\rangle\,,
\end{align}
where
\begin{align}
|\psi_\beta(0)\rangle
=
\frac{1}{\sqrt{Z(\beta)}}
\sum_j
e^{-\beta E_j/2}
|E_j\rangle_L\otimes|E_j\rangle_R\,.
\end{align}
If the state evolves with the Hamiltonian whose eigenvalues are $E_j$,
the survival amplitude takes the simple form
\begin{align}
S_\beta(t)
=
\frac{1}{Z(\beta)}
\sum_j
e^{-\beta E_j}e^{iE_jt}
=
\frac{Z(\beta-it)}{Z(\beta)}\, .
\label{eq:TFD_survival}
\end{align}
With this convention, the zeroth Krylov amplitude satisfies
\begin{align}
\psi_0(t)
=
S_\beta(t)^* .
\end{align}
The Hamiltonian moments are obtained by differentiating the survival
amplitude,
\begin{align}
\begin{split}
\mu_n
&\equiv
\langle\psi_\beta(0)|H^n|\psi_\beta(0)\rangle\,,
\\&=
\frac{1}{i^n}
\left.
\frac{d^n}{dt^n}S_\beta(t)
\right|_{t=0}=
\frac{1}{Z(\beta)}
\text{Tr}\!\left(e^{-\beta H}H^n\right)\,.
\label{5.5y}
\end{split}
\end{align}
The sequence $\{\mu_n\}$ determines the corresponding Lanczos coefficients through the standard moment--Lanczos construction. Equivalently, one may introduce the Hankel determinants,
\begin{align}
D_n
\equiv
\det\!\left[\mu_{i+j}\right]_{i,j=0}^{n-1}\,,
\qquad
D_0\equiv1\,,
\label{eq:Hankel_det}
\end{align}
in terms of which the off-diagonal Lanczos coefficients satisfy,
\begin{align}
b_n^2
=\frac{D_{n+1}D_{n-1}}{D_n^2}\,,
\qquad
n\geq1 \,.
\label{eq:Hankel_bn}
\end{align}
The diagonal coefficients $a_n$ can likewise be reconstructed from the
same moment sequence. In practice, below we determine both $a_n$ and $b_n$
numerically from the moments extracted from the semiclassical BTZ
partition function.

Having established the state Krylov construction, we now apply it to the
TFD state of the two-sided BTZ black hole. The analytic continuation of its
Euclidean partition function determines the survival amplitude
\eqref{eq:TFD_survival}, from which we obtain the moments, Lanczos
coefficients, and ultimately the Krylov spread complexity.

\subsection{Computation of spread complexity}\label{sub42}

We now apply the state Krylov construction to the TFD state
of the two-sided BTZ black hole. In the semiclassical large-$c$ regime, the
high-temperature partition function is dominated by the modular transform of the vacuum character,
\begin{align}
\begin{split}
Z(\tau,\bar\tau)
&\simeq
\chi_0\left(-\frac{1}{\tau}\right)
\bar\chi_0\left(-\frac{1}{\bar\tau}\right)\,,\\&=\frac{(1-q)(1-\bar{q})}{|\eta(\tau)|^2}(q\bar{q})^{\frac{-c}{24}}\Bigg|_{q=q_S=e^{-\frac{2\pi i}{\tau}}}
    \approx|q|^{\frac{-c}{12}}\Bigg|_{q=q_S=e^{-\frac{2\pi i}{\tau}}}=\exp\left(\frac{\pi^2c}{3\beta}\right)\,,
    \end{split}
\label{eq:BTZ_partition}
\end{align}
where $\beta$ denotes the inverse temperature.\footnote{The elliptic nome and its modular transform are
$q=e^{2\pi i\tau}$ and $q_S=e^{-2\pi i/\tau}$, respectively.} For the convention
$L=2\pi$ used throughout, the modular parameter of the non-rotating BTZ
boundary torus is
\begin{equation}
    \tau=\frac{i\beta}{2\pi}.
\end{equation}

Using \eqref{eq:TFD_survival}, the corresponding survival amplitude is
therefore given by,
\begin{align}
S_\beta(t)
=\frac{Z(\beta-it)}{Z(\beta)}
=\exp\left(
\frac{i\pi^2ct}{3\beta(\beta-it)}
\right)\,.
\label{eq:BTZ_survival}
\end{align}
The Hamiltonian moments introduced in \eqref{5.5y} are obtained by
differentiating the survival amplitude,
\begin{align}
\mu_n
&=
\frac{1}{i^n}
\left.
\frac{d^n}{dt^n}S_\beta(t)
\right|_{t=0}\,,
\nonumber\\
&=
n!\,\beta^{-n}\,
\mathbb L_n^{-1}
\left(
-\frac{\pi^2c}{3\beta}
\right)
\label{5.21w}
\end{align}
where $\mathbb L_n^{-1}$ denotes a generalized Laguerre polynomial. In
obtaining \eqref{5.21w}, we used
\begin{align}
\frac{d^n}{dx^n}
\left(x^\lambda e^{-a/x}\right)
=
(-1)^n n!\,
e^{-a/x}x^{\lambda-n}
\mathbb L_n^{(-\lambda-1)}
\left(\frac{a}{x}\right)\,.
\end{align}
The moments \eqref{5.21w} determine the Lanczos coefficients through the
moment--Lanczos construction reviewed in Sec.~(\ref{sub41}). The first few
coefficients are displayed in Table~(\ref{tab:lanczos}).

\begin{table}[t!]
\centering
\renewcommand{\arraystretch}{2.1}
\setlength{\tabcolsep}{10pt}
\rowcolors{2}{gray!5}{white}
\resizebox{\textwidth}{!}{%
\begin{tabular}{c c c}
\rowcolor{gray!12}
\textbf{$n$} & \textbf{$a_n$} & \textbf{$b_n$} \\
\toprule

$0$ & $\displaystyle \frac{c\pi^2}{3\beta^2}$ & $\displaystyle 0$ \\

$1$ & $\displaystyle \frac{c\pi^2+9\beta}{3\beta^2}$ &
$\displaystyle \pi\sqrt{\frac{2c}{3\beta^3}}$ \\

$2$ & $\displaystyle \frac{c\pi^2}{3\beta^2}+\frac{6}{\beta}-\frac{9}{4c\pi^2+9\beta}$ &
$\displaystyle \sqrt{\frac{4c\pi^2+9\beta}{3\beta^3}}$ \\

$3$ &
$\displaystyle \frac{c\pi^2}{3\beta^2}+\frac{9}{\beta}+\frac{9}{4c\pi^2+9\beta}
-\frac{18\left(8c^2\pi^4+45c\pi^2\beta+54\beta^2\right)}
{16c^3\pi^6+144c^2\pi^4\beta+405c\pi^2\beta^2+324\beta^3}$ &
$\displaystyle \frac{\sqrt{32c^3\pi^6+288c^2\pi^4\beta+810c\pi^2\beta^2+648\beta^3}}
{\beta^{3/2}\left(4c\pi^2+9\beta\right)}$ \\

\bottomrule
\end{tabular}%
}
\caption{The first few Lanczos coefficients $a_n$ and $b_n$ obtained from the moments.}
\label{tab:lanczos}
\end{table}

\begin{figure}[h!]
    \centering
\includegraphics[width=0.40\linewidth]{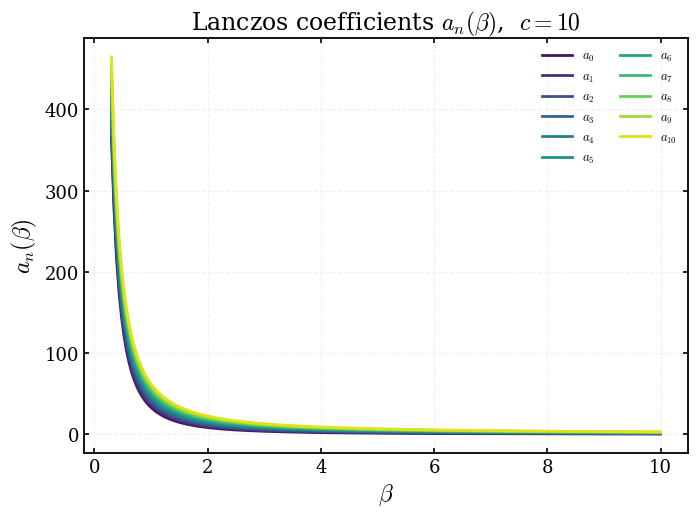}
\includegraphics[width=0.40\linewidth]{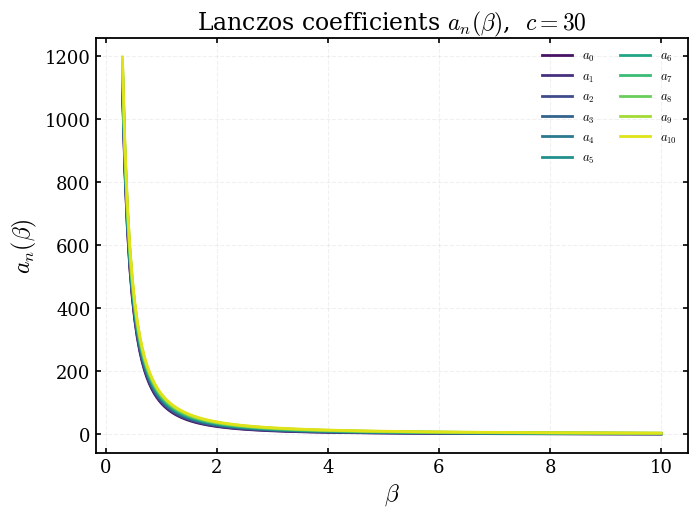}
\includegraphics[width=0.40\linewidth]{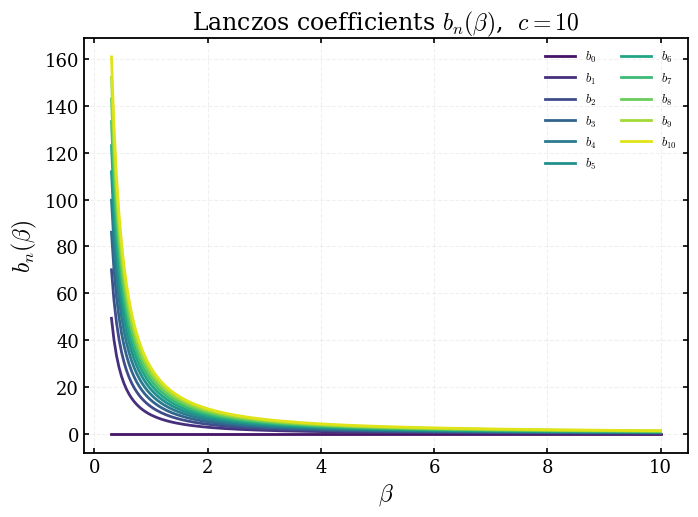}
\includegraphics[width=0.40\linewidth]{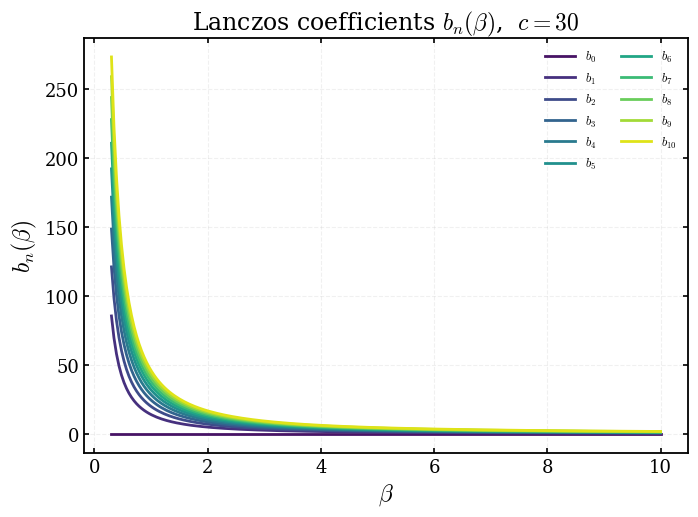}
    \caption{Dependence of the Lanczos coefficients $a_n$ and $b_n$ on
the inverse temperature $\beta$. The left column corresponds to
$c=10$, and the right column to $c=30$.}
    \label{fig:placeholder}
\end{figure}

Numerically, the Lanczos coefficients display a smooth dependence on the
Krylov index. Over the range accessible to our calculation, their behavior
is well described by
\begin{align}
b_n
\simeq
\alpha\sqrt{n(c_1n+d)},
\qquad
a_n
\simeq
a_\star+\gamma n,
\label{4.18u}
\end{align}
where $\alpha$, $\gamma$, $c_1$, and $d$ are determined by fitting the
Lanczos data at fixed $(c,\beta)$. The $n$-independent contribution
$a_\star$ produces only an overall phase in the Krylov amplitudes and will
be omitted below. Their numerical dependence on $\beta$ for representative values of the central charge is
shown in Fig.~(\ref{fig:placeholder}). 

At large Krylov index,
\begin{align}
b_n
=\alpha\sqrt{c_1}\,n
\left[
1+\frac{d}{2c_1n}
+\mathcal O\left(\frac{1}{n^2}\right)
\right],
\end{align}
so that the leading behavior is linear,
\begin{align}
b_n\simeq\alpha\sqrt{c_1}\,n\,,
\qquad
a_n\simeq\gamma n\,.
\label{eq:large_n_lanczos}
\end{align}
This approximately linear behavior is reminiscent of an
$\mathfrak{su}(1,1)$-type Krylov dynamics. However, as we discuss in
Appendix~\eqref{appE}, this resemblance must be interpreted with care:
the semiclassical BTZ moment problem is not
exactly an $\mathfrak{su}(1,1)$ chain, with deviations appearing at subleading order in $1/n$. In Appendix~\eqref{sec:krylov_fluctuations}, we further compare fluctuations of the
Krylov probability distribution with those of an exact $\mathfrak{su}(1,1)$ chain,
providing an independent diagnostic of the departure from exact
$\mathfrak{su}(1,1)$ dynamics.

\begin{figure}[t!]
    \centering
    \includegraphics[width=0.48\linewidth]{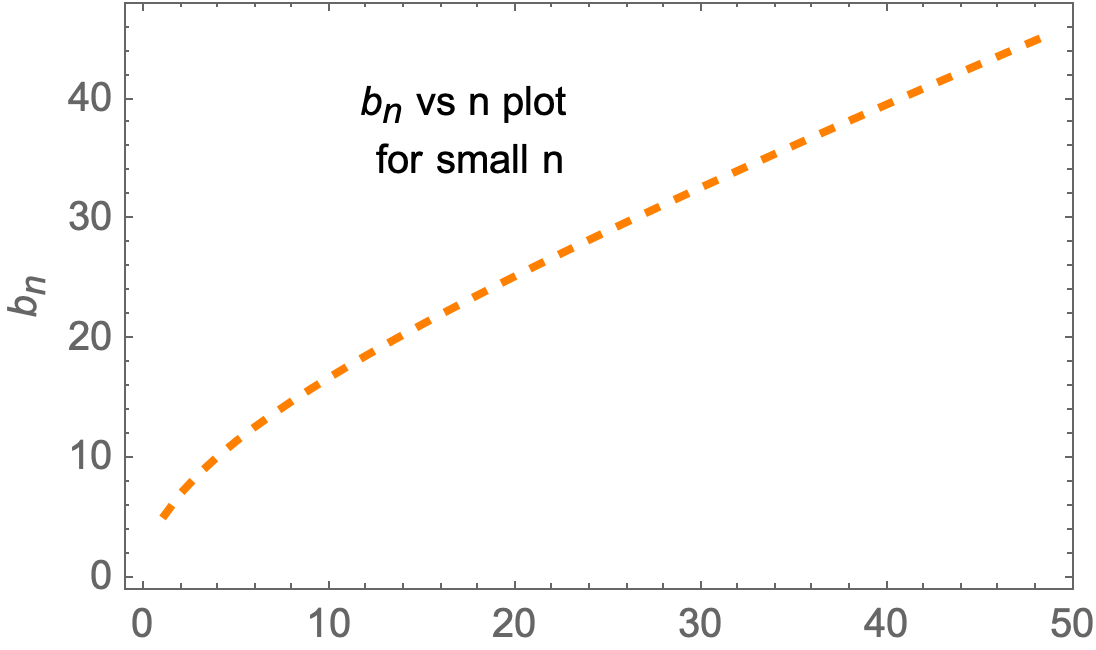}\,\,\,\,\,\,\,\,\,\,
    \includegraphics[width=0.45\linewidth]{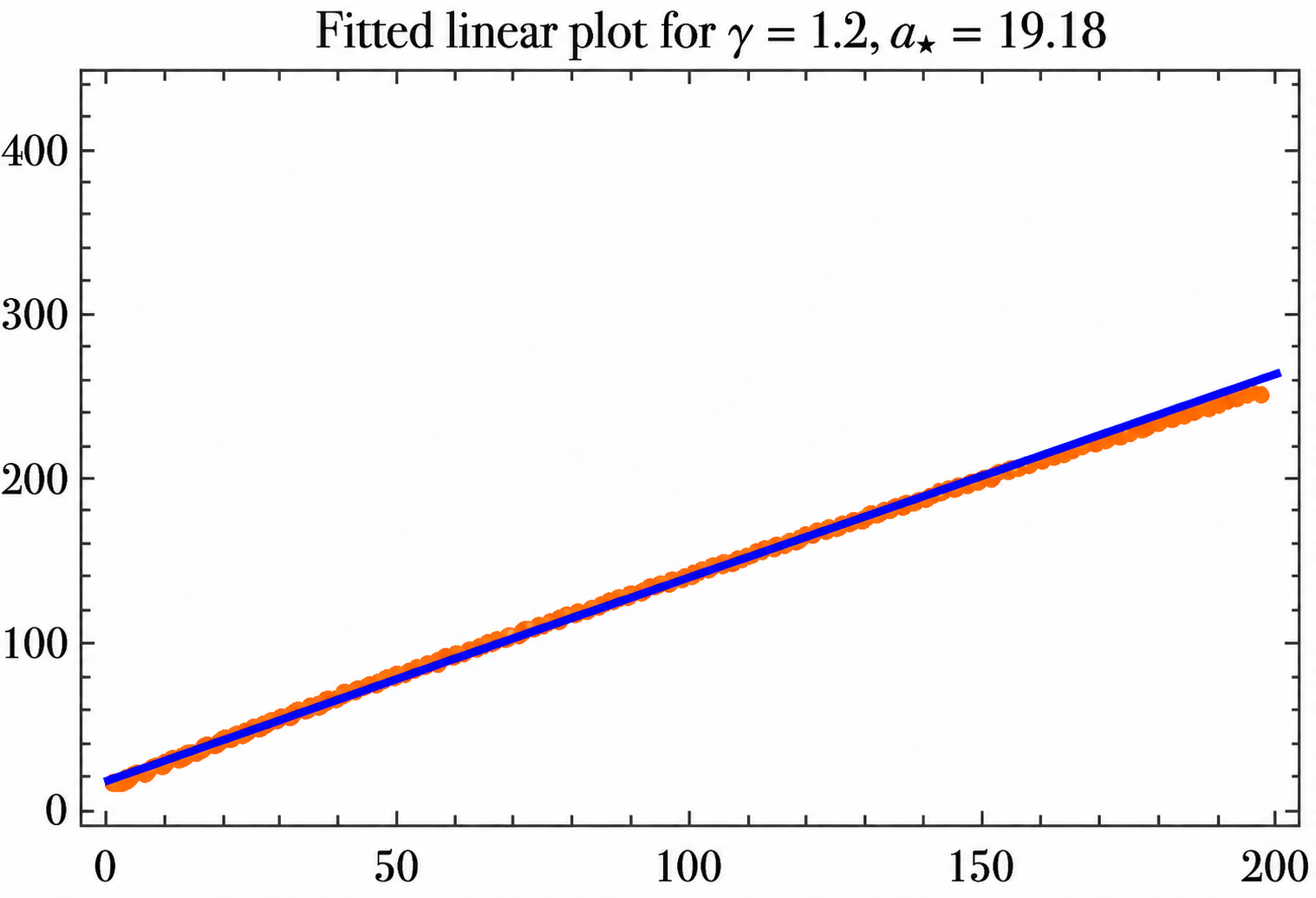}
\includegraphics[width=0.50\linewidth]{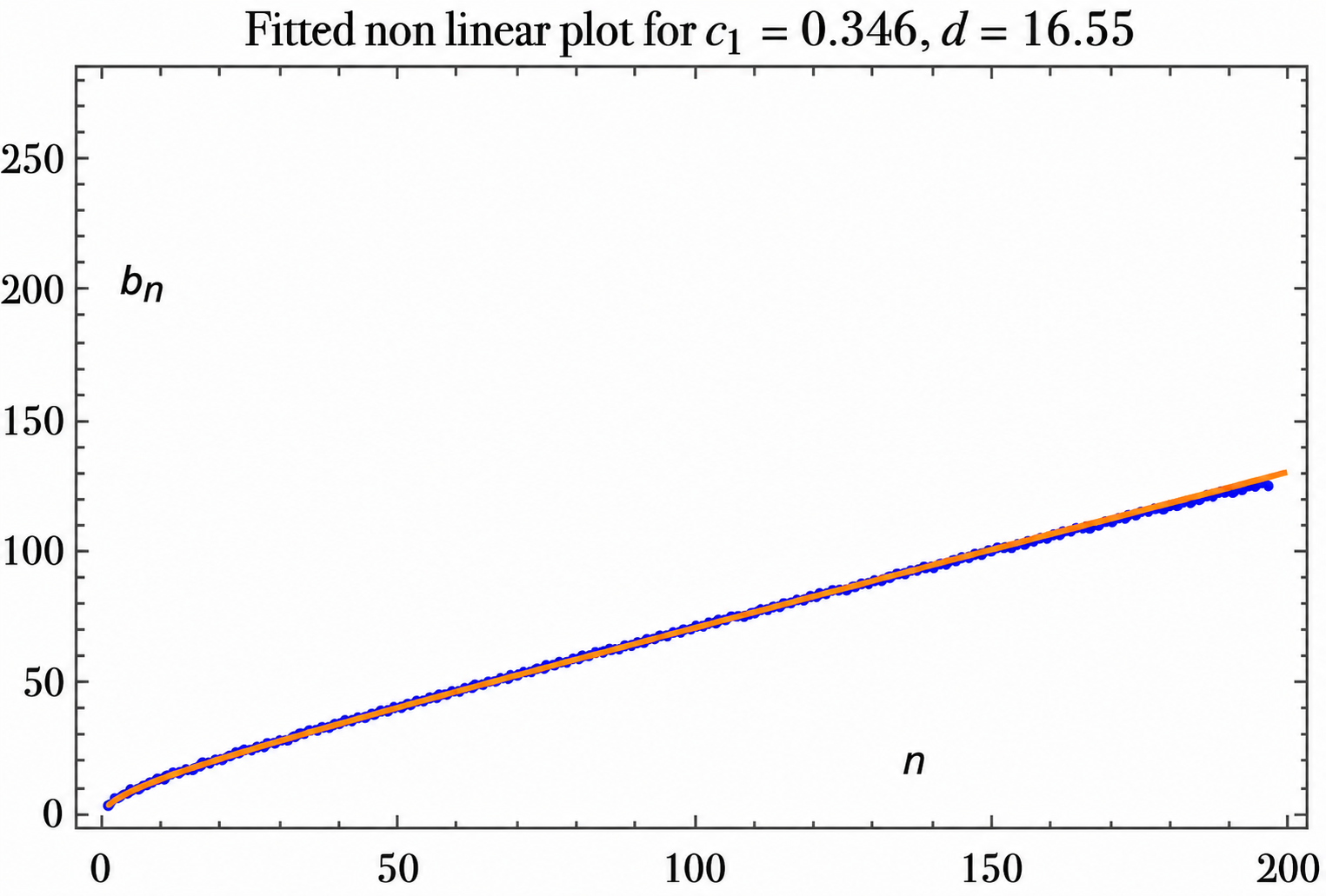}
\caption{Numerical Lanczos coefficients and representative fits to
Eq.~\eqref{4.18u}. The first panel shows $b_n$ at small Krylov index, while
the remaining panels display fits of $a_n$ and $b_n$ over larger ranges of
$n$. The numerics are truncated at $n=200$; extending the analysis to larger
$n$ requires increased precision to obtain reliable fits. The fitted
parameter values are indicated in each panel.}
    \label{fig:5}
\end{figure}

Over the finite range of Krylov indices used in the numerical fit, the
Lanczos coefficients are indeed well described by \eqref{4.18u}. Representative
fits are shown in Fig.~(\ref{fig:5}).
If such a fit is extrapolated to large $n$, the
leading coefficients take the form~\eqref{eq:large_n_lanczos} and the effective Krylov
recursion becomes
\begin{align}
i\partial_t\psi_n
\simeq
\gamma n\,\psi_n
+
\alpha\sqrt{c_1}
\left[
(n+1)\psi_{n+1}
+
n\psi_{n-1}
\right]\,.
\label{eq:effective-su-recursion}
\end{align}
This effective problem can be solved analytically using the generating
function method described in Appendix~\eqref{appC}. Introducing
\begin{align}
\omega
=
\frac{1}{2}
\sqrt{4\alpha^2c_1-\gamma^2}\,,
\label{eq:omega}
\end{align}
one finds, for the effective chain,
\begin{align}
\mathcal K_{\rm S}^{\rm eff}(t)
=
\frac{\alpha^2c_1}{\omega^2}
\sinh^2(\omega t)\,.
\label{eq:effective-spread}
\end{align}
For a noncritical fit satisfying $4\alpha^2c_1>\gamma^2$, this expression
would predict hyperbolic, and hence exponential, growth at sufficiently
large times.

This conclusion, however, relies on extending the finite-range fit
\eqref{4.18u} over the entire Krylov chain. As shown in detail in Appendix~\eqref{appE}, such an extrapolation is not consistent with the full
moment problem. In particular, the resulting zeroth Krylov amplitude does
not reproduce the survival amplitude $S_\beta(t)^*$ implied by the same
semiclassical partition function. This mismatch originates from replacing
the full finite-$n$ Lanczos sequence by its approximately linear form and
then extending this approximation down to $n=0$.

A direct analysis of the Lanczos coefficients generated by the
semiclassical moments reveals a different asymptotic structure. At large
Krylov index, our numerical results indicate
\begin{align}
a_n
\sim
\frac{2n}{\beta}\,,
\qquad
b_n
\sim
\frac{n}{\beta}\,.
\label{eq:true-asymptotic-Lanczos}
\end{align}
Thus, while the Lanczos chain becomes asymptotically
$\mathfrak{su}(1,1)$-like, its slopes approach the critical relation
\begin{align}
\gamma
=
2\alpha\sqrt{c_1}.
\label{eq:critical-main}
\end{align}
Equivalently, the effective parameter \eqref{eq:omega} approaches
$\omega=0$. The exponential behavior obtained from a noncritical
finite-range fit should therefore not be interpreted as the true
large-$n$ behavior of the semiclassical moment problem.

At the level of the effective $\mathfrak{su}(1,1)$ description, the critical limit is obtained smoothly from \eqref{eq:effective-spread},
\begin{align}
\lim_{\omega\to0}
\mathcal K_{\rm S}^{\rm eff}(t)
=\alpha^2c_1\,t^2.
\label{eq:critical-complexity}
\end{align}
The critical asymptotic slopes in \eqref{eq:true-asymptotic-Lanczos}
would therefore correspond to an approximately quadratic growth rather
than an exponential one.

This expectation is supported by a direct numerical evolution of the
Krylov chain constructed from the semiclassical moment problem. Using the
first $N=500$ Lanczos coefficients, we find that the spread complexity
clearly departs from the exponential behavior predicted by the noncritical
approximation and instead remains approximately quadratic over the
accessible time range. As shown in Fig.~(\ref{fig:7}) of Appendix~(\ref{appE}), this behavior is
consistent with the critical asymptotic relation
\eqref{eq:critical-main} and is strongly suggestive of a persistent quadratic regime at later times.

We therefore conclude that the Krylov spread complexity obtained from the
semiclassical BTZ partition function does not exhibit the linear growth
characteristic of the codimension-one bulk volume. Instead, the moment
problem generates an asymptotically critical
$\mathfrak{su}(1,1)$-like Krylov chain whose dynamics is compatible with
approximately quadratic growth. This provides a sharp contrast with the
operator Krylov complexity studied in Sec.~(\ref{secc5}), which develops the
linear late-time behavior characteristic of the BTZ interior.

\section{Conclusions and outlook}\label{sec5}

Motivated by the relation between quantum complexity and the growth of black hole interiors, in this work, we investigated how this connection extends beyond two-dimensional gravity. Focusing on the two-sided BTZ black hole, we developed a boundary description of the growth of the codimension-one surface entering the Complexity$=$Volume proposal and explored to what extent the same dynamics were encoded in different notions of Krylov complexity. Our main results may be summarized as follows:

\medskip
\noindent\textbf{Summary of results.}
Our first result was a boundary reconstruction of the growth of the BTZ interior.
In contrast to AdS$_2$/JT gravity, where the relevant codimension-one object is a
geodesic anchored at isolated boundary points \cite{Iliesiu:2021ari}, in AdS$_3$
the corresponding surface is anchored on spatial circles. This motivated us to
replace local boundary insertions by non-local operators obtained by smearing
primary operators along the spatial $S^1$. We chose a periodic Gaussian smearing
kernel, which reduces to a local insertion in the vanishing-width limit. We then
performed an additional averaging over appropriate segments of the Euclidean
thermal circle. We proposed that the resulting time-integrated two-point function
in the TFD state provides a boundary observable for the growth of the codimension-one
surface extending through the BTZ interior.

We then evaluated this observable using the Chern--Simons formulation of
three-dimensional gravity. Extending the Wilson-line construction to incorporate the spatial smearing
of the boundary insertions and the same Euclidean-time averaging, we expressed the observable in
terms of Chern--Simons group elements and found the characteristic linear growth
of the codimension-one BTZ surface at late times. This provides a nontrivial check of our proposed reconstruction and suggests that the relation between Wilson lines and
codimension-two observables, familiar from holographic entanglement entropy,
admits an extension to the codimension-one observables relevant to complexity.

In the second part of the paper, we asked whether this geometric growth was encoded in Krylov complexity.
For operator Krylov complexity, we considered the same spatially  smeared non-local operators
used in the reconstruction of the BTZ interior. From their two-point functions, we
extracted the weighted spectral measure and corresponding moments within a
microcanonical window. The resulting large-order moments implied that the Lanczos
coefficients $b_n$ approach a constant at large $n$, giving rise to linear late-time
operator Krylov complexity that mirrors the characteristic linear growth of the
codimension-one bulk observables. Although the overall coefficient may depend on
the smearing kernel, the robust feature is the common linear time dependence rather
than a universal equality of normalizations. This provides evidence that operator
growth can encode the expansion of the black hole interior beyond two-dimensional
gravity.

The picture was qualitatively different for Krylov spread complexity. Starting from
the semiclassical gravitational partition function of pure AdS$_3$ gravity, we
reconstructed the moment problem and the associated Lanczos dynamics of the TFD
state. Although the Lanczos coefficients become asymptotically
$\mathfrak{su}(1,1)$-like, the moment problem does not generate an exact
$\mathfrak{su}(1,1)$ chain. Instead, the asymptotic coefficients approach the
critical regime of the effective dynamics. Consistently, the spread complexity
remains approximately quadratic over the accessible time range, with the numerical
results strongly suggesting that this behavior persists at late times. We therefore
do not find a regime reproducing the characteristic linear growth of the bulk
volume.

Taken together, our results highlight a clear contrast between the two Krylov
constructions. In the semiclassical regime studied here, operator Krylov complexity
captures the characteristic linear growth of the BTZ interior, whereas the Krylov
spread complexity of the TFD state exhibits a qualitatively different, approximately
quadratic behavior. Whether quantum or non-perturbative corrections modify this
conclusion remains an important open question. More broadly, our results suggest
that the relation between Krylov complexity and black hole interior growth depends
sensitively on the choice of Krylov construction, with operator and state complexity
probing distinct aspects of the underlying dynamics.

\medskip
\noindent{\textbf{Future directions.}}
Our results suggest several directions for further investigation:
\begin{itemize}

\item\textbf{\textit{Non-perturbative completion in three-dimensional gravity:}}
A particularly important question is how the present semiclassical analysis is modified by non-perturbative effects. Unlike JT gravity, pure gravity in AdS$_3$ does not yet admit a comparably established matrix-model description, although substantial progress has been made through Virasoro TQFT, the Virasoro minimal string, and statistical
descriptions of CFT$_2$ data \cite{Collier:2023fwi,Collier:2023cyw,deBoer:2025oge}. Developing a sufficiently complete non-perturbative framework would make it possible to study how the boundary reconstruction proposed here is modified at exponentially late times and whether the operator and spread complexities eventually saturate. It would also be interesting to
identify the non-perturbative saddles responsible for such effects and determine their imprint on the Lanczos data.

\item\textbf{\textit{de Sitter and sine-dilaton analogs:}}
It would be interesting to extend our analysis to gravitational systems with de Sitter vacua. The complex Liouville string provides a particularly tractable framework in which sine-dilaton gravity admits both AdS$_2$ and dS$_2$ sectors and is related to a
two-matrix integral \cite{Collier:2024kmo}. This raises the possibility of studying both
perturbative and non-perturbative aspects of Krylov dynamics in a setting where the
gravitational interpretation differs substantially from the AdS black hole problem.
More broadly, it would be valuable to determine whether analogs of the relation between
operator growth and bulk geometry found here persist in de Sitter gravity, and how
horizon entropy, the absence of a timelike asymptotic boundary, and the different
global structure are encoded in Krylov space.

\item\textbf{\textit{Eigenstate thermalization for non-local operators:}}
Our construction naturally raises the question of whether, and in what sense, an
ETH-like description applies to the smeared non-local operators considered here.
Computing their matrix elements more directly in holographic CFTs could clarify how
the smearing kernel modifies the associated spectral measure and the large-$n$
behavior of the Lanczos coefficients. Such an analysis could also help determine which
features of the linear operator-complexity growth are universal and which depend on the
specific choice of extended operator, providing a useful probe of the interplay between
non-locality, scrambling, and quantum chaos.

\item\textbf{\textit{Long-time behavior of Krylov spread complexity:}}
Our semiclassical analysis of the TFD state yields approximately quadratic
growth over the accessible time range, with numerical results strongly suggesting
that this behavior persists at late times. Establishing the strict asymptotic behavior
and the effects of quantum and non-perturbative corrections, including eventual
saturation, remains an important open problem. It would also be interesting to
determine whether a linear regime analogous to that of the bulk volume can emerge
at later times. Its absence may instead indicate that Krylov spread complexity does
not generically track interior-volume growth beyond the special two-dimensional
examples where a direct correspondence is known. Resolving this question would
further sharpen the distinction between operator and state notions of Krylov
complexity.

\item\textbf{\textit{General smearing kernels and non-homogeneous operators:}}
Finally, our construction can be generalized by considering more general spatial
smearing profiles, including non-homogeneous kernels and varying the smearing scale
$\sigma$. This would allow us to investigate how the spectral measure, Lanczos
coefficients, and operator complexity depend on the detailed spatial support of the
boundary operator. More ambitiously, different choices of kernel could probe different
codimension-one bulk observables, potentially providing a concrete way to explore the
broader complexity $=$ anything framework from the boundary perspective.

\end{itemize}

\noindent
We hope to revisit some of these questions in future work.\\

\noindent
\textbf{Note Added:} While we were preparing this manuscript, a related work
\cite{Bhattacharya:2026ybq} appeared, studying Krylov spread complexity of the TFD
state from the semiclassical BTZ partition function. Both analyses find the universal
quadratic growth at early times and a departure from the linear growth characteristic
of the CV proposal. Using an early-time expansion together with Pad\'e continuation,
Ref.~\cite{Bhattacharya:2026ybq} finds behavior compatible with a return toward
quadratic growth in the pre-saturation regime, while emphasizing that the strict
late-time asymptotics is not established by the Pad\'e analysis alone. Our independent
reconstruction of the Lanczos chain from the semiclassical moment problem leads to a
consistent picture: the spread complexity remains approximately quadratic over the
accessible time range, with the numerical results strongly suggesting that this behavior
persists at late times. It would be interesting to understand in more detail the relation
between these complementary approaches. Independently, our analysis also considers
operator Krylov complexity for smeared non-local operators, for which we find a
late-time linear growth mirroring that of the codimension-one bulk surface.

\section*{Acknowledgments}
We would like to thank Rafael Carrasco, Saptaswa Ghosh, and Le-Chen Qu for useful discussions. AB is supported by the Core Research Grant (CRG/2023/001120) from the Anusandhan National Research Foundation, India. AB also acknowledges the associateship program of the Indian Academy of Sciences, Bengaluru, and support from the Indian Institute of Technology Gandhinagar and a generous donor through the Singheswari and Ram Krishna Jha Chair. JFP is supported by the ‘Atracción de Talento’ program of the Comunidad de Madrid under grant 2020-T1/TIC-20495, the Spanish Agencia Estatal de Investigación through grants CEX2025-001574-S, PID2021-123017NB-I00, and PID2024-156043NB-I00, funded by MCIN/AEI/10.13039/501100011033, and ERDF, EU. SP (PMRF ID: 1703278) is supported by the Prime Minister's Research Fellowship of the Government of India. We acknowledge the use of ChatGPT 5.6 Sol-Ultra for language editing and presentation of the manuscript.

\appendix

\section{Derivation of the two-point function}
\label{appAA}

The integral entering the two-point function is
\begin{align}
\label{eq:In-def}
I_n(t)
=
\int_{-2\pi R}^{2\pi R}
\frac{dx\,\exp\left(2\pi i \frac{n}{L}x\right)}
{\left[
\cosh\left(\frac{2\pi x}{\beta}\right)
+\cosh\left(\frac{2\pi t}{\beta}\right)
\right]^{\Delta}}\, .
\end{align}
Here $L$ denotes the circumference of the spatial circle, as introduced
in Sec.~\eqref{ssec21}; for the BTZ convention $\phi\sim\phi+2\pi$, one has
$L=2\pi$. 

For convenience, we define
\begin{equation}
\label{eq:alpha-def}
\alpha=\frac{2\pi}{\beta}\,,
\qquad
p=\Delta\,,
\end{equation}
and perform the change of variables
\begin{align}
\label{eq:u-change}
u=e^{\alpha x}
=
\exp\left(\frac{2\pi x}{\beta}\right).
\end{align}
This gives
\begin{equation}
dx
=
\frac{1}{\alpha}\frac{du}{u}
=
\frac{\beta}{2\pi}\frac{du}{u},
\qquad
\exp\left(2\pi i\frac{n}{L}x\right)
=
u^{i\beta \frac{n}{L}}.
\end{equation}
The integration limits become
\begin{equation}
u_{\pm}
=
\exp\left(
\pm\frac{4\pi^2R}{\beta}
\right).
\end{equation}

Using
\begin{equation}
\cosh(\alpha x)
=
\frac{1}{2}\left(u+\frac{1}{u}\right),
\qquad
\cosh(\alpha t)
=
\frac{1}{2}\left(e^{\alpha t}+e^{-\alpha t}\right),
\end{equation}
the denominator can be factorized as
\begin{align}
\cosh(\alpha x)+\cosh(\alpha t)
&=
\frac{1}{2u}
\left[
u^2+u\left(e^{\alpha t}+e^{-\alpha t}\right)+1
\right]
\nonumber\\
&=
\frac{1}{2u}
\left(u+e^{\alpha t}\right)
\left(u+e^{-\alpha t}\right).
\end{align}
It follows that we can combine,
\begin{align}
\left[
\cosh(\alpha x)+\cosh(\alpha t)
\right]^{-p}
=
2^p u^p
\left(u+e^{\alpha t}\right)^{-p}
\left(u+e^{-\alpha t}\right)^{-p}.
\end{align}

Substituting these expressions into \eqref{eq:In-def}, we obtain
\begin{equation}
\label{eq:In-u-integral}
I_n(t)
=
\frac{2^p\beta}{2\pi}
\int_{u_-}^{u_+}
du\,
u^{p+i\beta\frac{n}{L}-1}
\left(u+e^{\alpha t}\right)^{-p}
\left(u+e^{-\alpha t}\right)^{-p}.
\end{equation}
Equivalently, setting $p=\Delta$,
\begin{equation}
\label{eq:In-u-integral-Delta}
I_n(t)
=
\frac{2^{\Delta-1}\beta}{\pi}
\int_{u_-}^{u_+}
du\,
u^{\Delta+i\beta\frac{n}{L}-1}
\left(u+e^{\alpha t}\right)^{-\Delta}
\left(u+e^{-\alpha t}\right)^{-\Delta}.
\end{equation}

We now define
\begin{equation}
\label{eq:lambda-def}
\lambda
=
\Delta+i\beta\frac{n}{L}.
\end{equation}
The remaining integral is of Appell $F_1$ type. Using the identity
\begin{equation}
\label{eq:Appell-identity}
\int du\,
u^{\lambda-1}
(1+au)^{-p}
(1+bu)^{-p}
=
\frac{u^\lambda}{\lambda}
F_1\left(
\lambda;
p,p;
\lambda+1;
-au,-bu
\right),
\end{equation}
together with
\begin{align}
u+e^{\alpha t}
&=
e^{\alpha t}
\left(1+ue^{-\alpha t}\right),
\nonumber\\
u+e^{-\alpha t}
&=
e^{-\alpha t}
\left(1+ue^{\alpha t}\right),
\end{align}
we see that the overall factors $e^{-\alpha p t}$ and
$e^{+\alpha p t}$ cancel. An antiderivative is therefore
\begin{equation}
\Phi(u)
=
\frac{u^\lambda}{\lambda}
F_1\left(
\lambda;
\Delta,\Delta;
\lambda+1;
-ue^{-\alpha t},
-ue^{\alpha t}
\right).
\label{A.14r}
\end{equation}
Hence,
\begin{equation}
\label{eq:In-final-compact}
I_n(t)
=
\frac{2^{\Delta-1}\beta}{\pi}
\left[
\Phi(u_+)-\Phi(u_-)
\right],
\qquad
u_{\pm}
=
\exp\left(
\pm\frac{4\pi^2R}{\beta}
\right).
\end{equation}

Explicitly, the result can be written as
\begin{align}
\label{eq:In-final}
I_n(t)
=
\frac{2^{\Delta-1}\beta}{\pi}
\Bigg[
&\frac{u_+^{\,\Delta+i\beta \frac{n}{L}}}
{\Delta+i\beta \frac{n}{L}}
F_1\left(
\Delta+i\beta \frac{n}{L};
\Delta,\Delta;
\Delta+i\beta \frac{n}{L}+1;
-u_+e^{-2\pi t/\beta},
-u_+e^{2\pi t/\beta}
\right)
\nonumber\\
&-
\frac{u_-^{\,\Delta+i\beta \frac{n}{L}}}
{\Delta+i\beta \frac{n}{L}}
F_1\left(
\Delta+i\beta \frac{n}{L};
\Delta,\Delta;
\Delta+i\beta \frac{n}{L}+1;
-u_-e^{-2\pi t/\beta},
-u_-e^{2\pi t/\beta}
\right)
\Bigg]\,.
\end{align}

\section{Details of the time-integrated two-point function for $\Delta=2$}
\label{appB}

As illustrated in Fig.~(\ref{fig2}), the construction requires an additional
integration over Euclidean time. This integration is performed along the thermal
circle associated with the Euclidean BTZ geometry. After evaluating the Euclidean
integral, we analytically continue to Lorentzian time.

Writing $\phi_1'-\phi_2'=x$, the relevant integral takes the form
\begin{align}
\begin{split}
\label{eq:In-time-def}
I_n(t)
={}&
\underbrace{4\pi R}_{\text{from the $\phi_1'+\phi_2'$ integration}}
\int_{-\frac{\beta}{4}-\tau^E}^{\frac{\beta}{4}+\tau^E}
d\tau^E_1
\int_{-\frac{\beta}{4}-\tau^E}^{\frac{\beta}{4}+\tau^E}
d\tau^E_2
\int_{-2\pi R}^{2\pi R}
\frac{dx\,\exp\left(2\pi i \frac{n}{L}x\right)}
{\left[
\cosh\left(\frac{2\pi x}{\beta}\right)
+\cos\left(\frac{2\pi(\tau^E_1+\tau^E_2)}{\beta}\right)
\right]^2}
\\
={}&
-\frac{
4i\pi\beta^2R(\beta+4it)\,
\mathcal{N}(t,\beta)
}{
4\pi
\left(e^{\frac{8\pi^2R}{\beta}}-1\right)^2
\left(e^{\frac{8\pi t}{\beta}}-1\right)^2
},
\qquad n=0,\quad \Delta=2\,.
\end{split}
\end{align}
Here, the second line denotes the result after analytic continuation to Lorentzian
time. For compactness, we define
\begin{equation}
\label{eq:compact-expression}
\begin{gathered}
\mathcal{A}\equiv e^{\frac{4\pi^2R}{\beta}},
\qquad
\mathcal{B}\equiv e^{\frac{4\pi t}{\beta}},
\\[1mm]
\begin{aligned}
\mathcal{N}(t,\beta)
={}&
-(\mathcal{A}^2-1)(\mathcal{B}^2-1)
\left(
1+\mathcal{A}^2+\mathcal{B}^2
+4\mathcal{A}\mathcal{B}
+\mathcal{A}^2\mathcal{B}^2
\right)
\\
&+
2\log\left(
\frac{1-\mathcal{A}\mathcal{B}}
{\mathcal{A}-\mathcal{B}}
\right)
\Bigg[
\mathcal{B}^2(\mathcal{A}^2-1)^2
+
\left(
2\mathcal{A}^2+\mathcal{B}^2
-6\mathcal{A}^2\mathcal{B}^2
+\mathcal{A}^4\mathcal{B}^2
+2\mathcal{A}^2\mathcal{B}^4
\right)
\Bigg].
\end{aligned}
\end{gathered}
\end{equation}

The late-time behavior of the time-integrated correlator is therefore
\begin{align}
I_0(t)
\underset{t\to\infty}{\sim}
\frac{\beta^2Rt}{2\pi}
\left[
\coth\left(\frac{4\pi^2R}{\beta}\right)
-\frac{4\pi^2R}{\beta}
\csch^2\left(\frac{4\pi^2R}{\beta}\right)
\right].
\end{align}
Thus, the integrated correlator exhibits the linear late-time behavior used in the main text.

\section{Computation of correlators in the TFD state}
\label{appD}

In this appendix, we briefly review the construction of bulk-local states in
AdS$_3$ and its extension to the BTZ geometry dual to the thermofield double
(TFD) state. Our main purpose is to obtain the two-sided thermal correlator
used in Sec.~(\ref{3.2rr}).

We begin with the extrapolate dictionary, which relates a bulk field near the
asymptotic boundary to a primary operator in the dual CFT. At the level of
states, this implies \cite{Goto:2016wme,Goto:2017olq},
\begin{align}
\lim_{\rho\to\infty} e^{\rho\Delta}
|\Phi_\alpha(\rho,\vec{x})\rangle
=
\lim_{y\to0} y^{-\Delta}
|\Phi_\alpha(y,\vec{x})\rangle
=
\mathcal{O}_\alpha(\vec{x})|0\rangle ,
\label{D.1w}
\end{align}
where $\Delta$ is the conformal dimension of the primary operator
$\mathcal{O}_\alpha$, while $(\rho,\vec{x})$ and $(y,\vec{x})$ denote bulk
coordinates in global and Poincar\'e AdS, respectively.

The same construction can be adapted locally to the BTZ geometry. In this
case, the bulk-local state can be represented as an infinite linear
combination of descendants acting on the TFD state,
\begin{align}
\bigl|\Phi_{\alpha}(\rho,\varphi,t)\bigr\rangle_{\mathrm{BTZ}}
\equiv
\sum_{k=0}^{\infty}
(-1)^k
\frac{\Gamma(\Delta)}{k!\,\Gamma(k+\Delta)}
\left(L_{-1}^{x}\right)^k
\left(\bar L_{-1}^{x}\right)^k
\mathcal{O}_\alpha(\varphi,\gamma)
\bigl|\Psi_{\mathrm{TFD}}\bigr\rangle ,
\label{3.2w}
\end{align}
where $\gamma=t_r-i\tau_{r,0}$. The generators
$L_{-1}^{x}$ and $\bar L_{-1}^{x}$ are defined locally through the conformal
map relating the boundary insertion point to the corresponding bulk point
\cite{Goto:2017olq}. Their repeated action provides the localizing combination of descendants required to reconstruct the bulk excitation. The TFD state describing the two-sided BTZ black hole is
\begin{equation}
\bigl|\Psi_{\mathrm{TFD}}\bigr\rangle
=
\frac{1}{\sqrt{Z(\beta)}}
\sum_E e^{-\beta E/2}
|E\rangle_L\otimes|E\rangle_R ,
\label{TFDstate}
\end{equation}
where $\beta$ is the inverse temperature. In the conventions used throughout
this work, the spatial coordinate is periodically identified as
$\varphi\sim\varphi+2\pi$. The Euclidean boundary of BTZ is therefore a
torus, whereas the corresponding Rindler-AdS construction is naturally
defined on the noncompact covering space. Since BTZ is locally AdS$_3$, its
bulk-local correlators can be obtained from the Rindler-AdS correlator,
supplemented by the appropriate spatial identifications.

Locally, the relation between the coordinates on the two descriptions may be
written as \cite{Goto:2017olq}
\begin{equation}
\tanh\left(\frac{\varphi\pm t}{2}\right)
=
\tan\left(\frac{\varphi'\pm t'}{2}\right),
\qquad
z=e^{\tau'+i\varphi'},
\qquad
\bar z=e^{\tau'-i\varphi'},
\label{localmap}
\end{equation}
where $\tau'=it'$. This map is only local, as expected from the different
global topologies of the Rindler-AdS and BTZ boundaries, but this is
sufficient for the construction of bulk-local states.

At leading order in the semiclassical large-$c$ limit, the two-point
function of bulk-local operators reproduces the Rindler-AdS bulk
propagator,
\begin{align}
\langle\Psi_{\mathrm{TFD}}|
\Phi(\rho,\varphi,t)
\Phi(\rho',\varphi',t')
|\Psi_{\mathrm{TFD}}\rangle
&=
\mathcal{G}_{\mathrm{Rindler}}
(\rho,\varphi,t;\rho',\varphi',t'),
\end{align}
with
\begin{align}
\mathcal{G}_{\mathrm{Rindler}}
=
\frac{1}{
2\sqrt{\sigma^2-1}
\left(\sigma+\sqrt{\sigma^2-1}\right)^{\Delta-1}} .
\label{C.13r}
\end{align}
Here $\sigma$ is the AdS-invariant distance. For two points in the same
exterior Rindler wedge, it takes the form
\begin{align}
\sigma(x_r|x'_r)
&=
\cosh\rho_r\cosh\rho'_r\cosh\Delta\varphi
-
\sinh\rho_r\sinh\rho'_r\cosh\Delta t
\nonumber\\
&=
\frac{1}{r_+^2}
\left[
rr'\cosh\Delta\varphi
-
\sqrt{r^2-r_+^2}\sqrt{r'^2-r_+^2}\cosh\Delta t
\right],
\label{D.sigma}
\end{align}
where $\Delta\varphi=\varphi-\varphi'$ and
$\Delta t=t-t'$. Correlators involving points in different exterior wedges
are obtained by the corresponding analytic continuation. In particular, for
insertions on opposite boundaries of the eternal black hole, this
continuation converts the dependence on the time difference into a
dependence on $t_L+t_R$.

The compact BTZ geometry is obtained by imposing the periodic
identification of the spatial coordinate. Equivalently, the corresponding
propagator can be represented as a sum over spatial images,
\begin{align}
\mathcal{G}_{\mathrm{BTZ}}
=
\sum_{n,m}
\mathcal{G}_{\mathrm{Rindler}}
\left(
\rho,\varphi+2\pi m,t;
\rho',\varphi'+2\pi n,t'
\right). \label{c.8}
\end{align}
In the high-temperature regime considered in the main text, the leading
Rindler contribution gives the local two-sided thermal correlator. Taking
the asymptotic boundary limit and analytically continuing the two insertions
to opposite boundaries yields
\begin{align}
\left\langle
\mathcal{O}^{L}(w_1)\mathcal{O}^{R}(w_2)
\right\rangle_{\beta}
=
\left(\frac{\pi}{\beta}\right)^{2\Delta}
\left[
\frac{N}{\epsilon^2}
\cosh\left(\frac{2\pi}{\beta}(t_L+t_R)\right)
+
\frac{N}{\epsilon^2}
\cosh\left(\frac{2\pi}{\beta}
(\varphi_1-\varphi_2)\right)
\right]^{-\Delta}.
\label{D.TFDcorr}
\end{align}
This is precisely the local thermal correlator used in \eqref{eq38app}.
Smearing the two boundary insertions with the kernel introduced in
Sec.~(\ref{ssec21}) then gives the extended correlator entering the computation of
operator Krylov complexity.

\section{Derivation of $\phi_n(t)$ from the generating function}
\label{appC}

We consider the recursion relation,
\begin{align}
i\dot{\phi}_n(t)
=\alpha\Big[
\sqrt{(n+1)\big(c_1(n+1)+d\big)}\,
e^{-i\gamma t}\phi_{n+1}(t)
+
\sqrt{n(c_1n+d)}\,
e^{i\gamma t}\phi_{n-1}(t)
\Big]\,.
\label{eqA.1}
\end{align}
We are interested in the large-$n$ regime, where the Lanczos coefficients
admit the expansion
\begin{align}
\sqrt{n(c_1n+d)}
=
\sqrt{c_1}\,n
\left[
1+\frac{d}{2c_1n}
+\mathcal O\left(\frac{1}{n^2}\right)
\right].
\end{align}
At leading order in the large-$n$ expansion, we therefore approximate the
Lanczos coefficients by
\begin{align}
\sqrt{n(c_1n+d)}
\simeq
\sqrt{c_1}\,n\,,
\end{align}
and solve the resulting effective recursion relation,
\begin{align}
i\dot{\phi}_n(t)
\simeq
\alpha\sqrt{c_1}
\left[
(n+1)e^{-i\gamma t}\phi_{n+1}(t)
+
n e^{i\gamma t}\phi_{n-1}(t)
\right].
\label{eq:large-n-recursion}
\end{align}

The explicit time-dependent phases can be removed by introducing
\begin{equation}
\phi_n(t)=e^{in\gamma t}\Psi_n(t)\,.
\end{equation}
Substituting this ansatz into \eqref{eq:large-n-recursion}, we obtain
\begin{equation}
i\dot{\Psi}_n(t)
=
\alpha\sqrt{c_1}
\Big[
(n+1)\Psi_{n+1}(t)
+n\Psi_{n-1}(t)
\Big]
+n\gamma\Psi_n(t)\,,
\label{A.15t}
\end{equation}
with $\Psi_{-1}(t)=0$. We impose the standard initial condition
\begin{align}
\Psi_n(0)=\delta_{n0}\,.
\end{align}

To solve \eqref{A.15t}, we introduce the generating function
\begin{align}
\mathbb G(z,t)
=
\sum_{n=0}^{\infty}\Psi_n(t)z^n\,.
\label{eq:G-def}
\end{align}
Multiplying \eqref{A.15t} by $z^n$ and summing over $n$, we find
\begin{equation}
i\partial_t\mathbb G(z,t)
=
\left[
\alpha\sqrt{c_1}(1+z^2)+\gamma z
\right]
\partial_z\mathbb G(z,t)
+
\alpha\sqrt{c_1}\,z\,\mathbb G(z,t)\,.
\label{A.17y}
\end{equation}

\subsection{Method of characteristics}

Equation~\eqref{A.17y} is a first-order PDE and can be solved using the
method of characteristics. Along a characteristic $z=z(t)$,
\begin{align}
\frac{d\mathbb G(z(t),t)}{dt}
=
\partial_t\mathbb G
+
\dot z\,\partial_z\mathbb G\,.
\label{A.18y}
\end{align}
Choosing
\begin{align}
\frac{dz(t)}{dt}
=
i\left[
\alpha\sqrt{c_1}\big(1+z(t)^2\big)
+\gamma z(t)
\right]\,,
\qquad
z(0)=z_0\,,
\label{A.17u}
\end{align}
\eqref{A.17y} reduces along the characteristic flow to,
\begin{align}
\frac{d\mathbb G}{dt}
=
-i\alpha\sqrt{c_1}\,
z(t)\mathbb G(z(t),t)\,.
\label{A.20i}
\end{align}
The characteristic equation~\eqref{A.17u} is of Riccati type. It can be
linearized through the substitution,
\begin{align}
z(t)
=
-\frac{1}{i\alpha\sqrt{c_1}}
\frac{\dot u(t)}{u(t)}\,,
\label{eq:riccati-substitution}
\end{align}
which gives,
\begin{align}
\ddot u-i\gamma\dot u-\alpha^2c_1u=0\,.
\label{B.17p}
\end{align}
Using \eqref{eq:riccati-substitution} in \eqref{A.20i}, we also obtain
\begin{align}
\mathbb G(z(t),t)
=
\mathbb G_0(z_0)
\frac{u(t)}{u(0)}\,,
\label{A.19y}
\end{align}
where $\mathbb G_0(z)=\mathbb G(z,0)\,.$
\subsection{Solution of the characteristic equation}
As Eq.~\eqref{B.17p} is a second-order differential equation, it can be cast as two coupled first-order ordinary differential equations which can be written as,
\begin{align}
\mathbf v(t)
=
\begin{pmatrix}
u(t)\\
\dot u(t)
\end{pmatrix},
\end{align}
so that
\begin{align}
\frac{d\mathbf v}{dt}
=
K\mathbf v,
\qquad
K=
\begin{pmatrix}
0 & 1\\
\alpha^2c_1 & i\gamma
\end{pmatrix}.
\end{align}
The solution can be written as,
\begin{align}
\mathbf v(t)
=M(t)\mathbf v(0)\,,
\qquad
M(t)=e^{Kt}.
\end{align}
The eigenvalues of $K$ are
\begin{align}
\lambda_{\pm}
=
\frac{i\gamma}{2}\pm\omega,
\qquad
\omega
\equiv
\frac{1}{2}\sqrt{4\alpha^2c_1-\gamma^2}\,.
\label{eq:omega-def}
\end{align}
Using the standard expression for the exponential of a $2\times2$ matrix,
we find
\begin{align}
M(t)
=
e^{\frac{i\gamma t}{2}}
\left[
\cosh(\omega t)\,I
+
\frac{\sinh(\omega t)}{\omega}
\left(
K-\frac{i\gamma}{2}I
\right)
\right],
\end{align}
or explicitly
\begin{align}
M(t)
=
e^{\frac{i\gamma t}{2}}
\begin{pmatrix}
\cosh(\omega t)
-\frac{i\gamma}{2\omega}\sinh(\omega t)
&
\frac{1}{\omega}\sinh(\omega t)
\\[6pt]
\frac{\alpha^2c_1}{\omega}\sinh(\omega t)
&
\cosh(\omega t)
+\frac{i\gamma}{2\omega}\sinh(\omega t)
\end{pmatrix}.
\label{A.33y}
\end{align}
Notice that
\begin{align}
\det M(t)=e^{i\gamma t},
\end{align}
so that $M(t)\in GL(2,\mathbb C)$. 
Below, we define the components of the matrix as follows,\\
\begin{align}
\begin{split}
\label{C.23r}
&M_{11}(t)=
e^{\frac{i\gamma t}{2}}
\left[
\cosh(\omega t)
-\frac{i\gamma}{2\omega}\sinh(\omega t)
\right],\,
\,\,M_{12}(t)=\,
e^{\frac{i\gamma t}{2}}
\frac{\sinh(\omega t)}{\omega}\,.
\end{split}
\end{align}After extracting the overall factor
$e^{i\gamma t/2}$, the remaining matrix has unit determinant. At $t=0$, \eqref{eq:riccati-substitution} implies
\begin{align}
z_0
=-\frac{1}{i\alpha\sqrt{c_1}}
\frac{\dot u(0)}{u(0)}
\qquad\Longrightarrow\qquad
\dot u(0)
=
-i\alpha\sqrt{c_1}\,z_0u(0)\,.
\end{align}
It follows that
\begin{align}
\frac{u(t)}{u(0)}
=
M_{11}(t)
-i\alpha\sqrt{c_1}\,M_{12}(t)z_0\,.
\label{A.40i}
\end{align}

We can now use the solution of the characteristic flow to determine the
generating function and extract the Lanczos amplitudes. For the initial
condition $\Psi_n(0)=\delta_{n0}$, the initial generating function is simply
\begin{align}
\mathbb G(z,0)=1\,.
\end{align}
Therefore, using the ratio in \eqref{A.40i} and inverting the flow obtained from solving the differential equation of $z$ is given by,
\begin{align}
  z_0{[z,t]}=  \frac{2 \omega  z+i \tanh (t \omega ) \left(2 \alpha  \sqrt{c_1}+\gamma  z\right)}{2 \omega -i \tanh (t \omega ) \left(2 \alpha  \sqrt{c_1} z+\gamma \right)}\,.\label{A.41w}
\end{align}
Hence, the generating function \eqref{A.19y} can be written as,
\begin{align}
    \mathbb{G}(z,t)=\mathbb{G}_0(z_0)\textcolor{black}{\frac{u(t)}{u(0)}}\,.
\end{align}
Expressing the characteristic solution in terms of the final coordinate $z$, the generating function takes the compact form,
\begin{align}
\mathbb G(z,t)
=
\frac{f(t)}{1-q(t)z}\,,
\label{eq:G-final}
\end{align}
where
\begin{align}
q(t)
&=-\frac{
2i\alpha\sqrt{c_1}\,\sinh(\omega t)}{
2\omega\cosh(\omega t)
+i\gamma\sinh(\omega t)
},
\,\,\,\,f(t)=
\frac{
2\omega\,e^{i\gamma t/2}
}{
2\omega\cosh(\omega t)
+i\gamma\sinh(\omega t)
}\,.
\label{eq:f-final}
\end{align}
Expanding \eqref{eq:G-final} in powers of $z$,
\begin{align}
\mathbb G(z,t)
=
f(t)
\sum_{n=0}^{\infty}
q(t)^n z^n\,,
\end{align}
we obtain\footnote{ using the relation,\begin{align}
 \frac{d^n}{dz^n}\Bigg(\frac{az+b}{cz+d}\Bigg)\bigg|_{z=0}= \frac{n! \left(-\frac{c}{d}\right)^n (b c-a d)}{c d}
\end{align}} 
\begin{align}
\Psi_n(t)
=\frac{
2\omega\,e^{i\gamma t/2}
}{
2\omega\cosh(\omega t)
+i\gamma\sinh(\omega t)
}
\left[
-\frac{
2i\alpha\sqrt{c_1}\,\sinh(\omega t)
}{
2\omega\cosh(\omega t)
+i\gamma\sinh(\omega t)
}
\right]^n.
\label{eq:Psi-final}
\end{align}
The original Lanczos amplitudes are therefore
\begin{align}
\phi_n(t)
=
e^{in\gamma t}
\frac{
2\omega\,e^{i\gamma t/2}
}{
2\omega\cosh(\omega t)
+i\gamma\sinh(\omega t)
}
\left[
-\frac{
2i\alpha\sqrt{c_1}\,\sinh(\omega t)
}{
2\omega\cosh(\omega t)
+i\gamma\sinh(\omega t)
}
\right]^n
\label{eq:phi-final}
\end{align}
with $\omega$ defined in \eqref{eq:omega-def}. The solution satisfies
\begin{align}
\phi_n(0)=\delta_{n0}\,,
\end{align}
as required. Moreover, for real $\omega$ one can explicitly verify that
\begin{align}
\sum_{n=0}^{\infty}
|\phi_n(t)|^2
=
1\,.
\end{align}

We emphasize that \eqref{eq:phi-final} solves the effective recursion
relation obtained from the leading large-$n$ behavior of the Lanczos
coefficients, rather than the full finite-$n$ recursion in
\eqref{eqA.1}. In particular, the parameter $d$ enters only through
subleading corrections in the large-$n$ expansion and is therefore absent
from the leading solution above. For
\begin{align}
4\alpha^2c_1<\gamma^2,
\end{align}
$\omega$ becomes imaginary and the hyperbolic functions analytically continue
to oscillatory functions. The critical case
$4\alpha^2c_1=\gamma^2$ is obtained smoothly by taking the limit
$\omega\to0$ in Eqs.~\eqref{eq:f-final}--\eqref{eq:phi-final}.

\section{Moment problem associated with the semiclassical partition function}
\label{appE}

In this appendix, we examine more carefully the moment problem associated
with the semiclassical BTZ partition function. The large-$n$ behavior of
the Lanczos coefficients found in Sec.~\eqref{sub42} suggests a structure reminiscent
of $\mathfrak{su}(1,1)$ dynamics. However, this identification cannot be
exact. A first indication is that the zeroth Krylov amplitude obtained by solving
the large-$n$ effective recursion in Appendix~\eqref{appC} does not reproduce
the survival amplitude implied by the same semiclassical partition function,
\begin{align}
\psi_0(t)=S_\beta(t)^*\,.
\label{eq:survival-boundary}
\end{align}
As we show below, this discrepancy does not signal an inconsistency in the
choice of initial state. Rather, it results from extrapolating the
asymptotic large-$n$ Lanczos coefficients down to the boundary of the
Krylov chain at $n=0$.

\paragraph{Moments and Lanczos coefficients of the semiclassical saddle.}

To make this point explicit, consider the semiclassical partition function
\begin{align}
Z(\beta)
=
\exp\left(\frac{\pi^2c}{3\beta}\right)\,.
\end{align}
It is convenient to introduce
\begin{align}
\label{E.3r}
\lambda
\equiv
\frac{\pi^2c}{3\beta}\,,
\qquad
x\equiv\beta H \,,
\end{align}
and work with the dimensionless moments
\begin{align}
\widetilde\mu_n
\equiv
\langle x^n\rangle_\beta
=
\beta^n\mu_n \,.
\end{align}
Using the survival amplitude associated with $Z(\beta)$, the first
few moments are
\begin{align}
\widetilde\mu_0 &= 1\,,
\nonumber\\
\widetilde\mu_1 &= \lambda\,,
\nonumber\\
\widetilde\mu_2 &= \lambda^2+2\lambda\,,
\nonumber\\
\widetilde\mu_3 &= \lambda^3+6\lambda^2+6\lambda\,,
\nonumber\\
\widetilde\mu_4
&=
\lambda^4+12\lambda^3+36\lambda^2+24\lambda\,,
\nonumber\\
\widetilde\mu_5
&=
\lambda^5+20\lambda^4+120\lambda^3
+240\lambda^2+120\lambda\,.
\label{eq:first-moments}
\end{align}
Using these moments, the first few Krylov amplitudes can be obtained explicitly.
Defining
\begin{align}
q\equiv\frac{\beta}{\beta+it},
\end{align}
one finds
\begin{align}
\begin{split}
\psi_0(t)
&=S_\beta(t)^*,\\
\psi_1(t)
&=\sqrt{\frac{\lambda}{2}}\left(q^2-1\right)S_\beta(t)^*,\\
\psi_2(t)
&=\sqrt{\frac{\lambda}{2(4\lambda+3)}}(q-1)^2
\left[\lambda(q+1)^2+2q+1\right]S_\beta(t)^*,\\
&\hspace{2mm}\vdots
\end{split}
\end{align}
These amplitudes satisfy $\psi_n(0)=\delta_{n0}$, as required.

Let $p_n(x)$ denote the orthonormal polynomials associated with the moment sequence \eqref{eq:first-moments}. They satisfy the three-term recurrence relation
\begin{align}
x p_n(x)
=\widetilde b_{n+1}p_{n+1}(x)
+\widetilde a_n p_n(x)
+\widetilde b_n p_{n-1}(x)\,,
\label{eq:orthogonal-recursion}
\end{align}
with $\widetilde b_0=0$. The physical Lanczos coefficients are related to
the dimensionless ones by
\begin{align}
a_n=\frac{\widetilde a_n}{\beta}\,,
\qquad
b_n=\frac{\widetilde b_n}{\beta}\,.
\label{eq:physical-coefficients}
\end{align}
Applying the Lanczos procedure directly to the semiclassical moments
\eqref{eq:first-moments} gives
\begin{align}
\widetilde a_0&=\lambda\,,
&
\widetilde b_1^2&=2\lambda\,,
\nonumber\\
\widetilde a_1&=\lambda+3\,,
&
\widetilde b_2^2&=4\lambda+3\,.
\label{eq:second-lanczos}
\end{align}
At the next orders, one finds
\begin{align}
\widetilde a_2
&=
\frac{4\lambda^2+27\lambda+15}{4\lambda+3}
=
\lambda+6-\frac{3}{4\lambda+3}\,,
\nonumber\\
\widetilde b_3^2
&=
6\,
\frac{
16\lambda^3+48\lambda^2+45\lambda+12
}{
(4\lambda+3)^2
}\,,
\label{eq:b3}
\end{align}
and
\begin{align}
\widetilde a_3
=
\frac{
64\lambda^5
+816\lambda^4
+2340\lambda^3
+2739\lambda^2
+1404\lambda
+252
}{
(4\lambda+3)
\left(
16\lambda^3
+48\lambda^2
+45\lambda
+12
\right)
}\,.
\label{eq:a3}
\end{align}
These expressions already show that the diagonal coefficients
$\widetilde a_n$ and the squared off-diagonal coefficients
$\widetilde b_n^2$ are nontrivial rational functions of $\lambda$.
In particular, the Lanczos coefficients do not coincide exactly with
those of an $\mathfrak{su}(1,1)$ chain.

\paragraph{Comparison with an exact $\mathfrak{su}(1,1)$ chain.}
\label{sec:su11}

To quantify this difference, consider an exactly solvable
$\mathfrak{su}(1,1)$ Hamiltonian
\begin{align}
\mathbf H_{\rm SU}
=
\Gamma K_0
+
\kappa(K_++K_-)
+
\delta \,.
\label{eq:su-hamiltonian}
\end{align}
For a positive discrete-series representation,
\begin{align}
K_0|k,n\rangle=(n+k)|k,n\rangle\,,
\end{align}
and the corresponding Lanczos coefficients take the form
\begin{align}
a_n^{\rm SU}
&=
\Gamma(n+k)+\delta\,,
\nonumber\\
b_n^{\rm SU}
&=
\kappa\sqrt{n(n+2k-1)}\,.
\label{eq:su-bn}
\end{align}
In particular, the diagonal coefficients are exactly linear in $n$ and
therefore satisfy
\begin{align}
a_{n+2}^{\rm SU}
-2a_{n+1}^{\rm SU}
+a_n^{\rm SU}
=0\,.
\label{eq:su-linearity-condition}
\end{align}
The coefficients generated by the semiclassical partition function
instead give
\begin{align}
\widetilde a_2
-2\widetilde a_1
+\widetilde a_0
&=
\left(
\lambda+6-\frac{3}{4\lambda+3}
\right)
-2(\lambda+3)
+\lambda
\nonumber\\
&=
-\frac{3}{4\lambda+3}
\neq0\,.
\label{eq:failure-su}
\end{align}
Thus, the failure of exact $\mathfrak{su}(1,1)$ dynamics is already visible
in the first few diagonal Lanczos coefficients.

The same conclusion follows from the off-diagonal coefficients. As discussed
below, the large-$n$ behavior fixes the dimensionless asymptotic hopping
coefficient to $\kappa=1$. With this normalization,
$\widetilde b_1^2=2\lambda$ would imply $k=\lambda$. An exact
$\mathfrak{su}(1,1)$ chain would then predict,
\begin{align}
\left(\widetilde b_2^{\rm SU}\right)^2
=
4\lambda+2\,,
\end{align}
whereas the semiclassical moment problem gives
\begin{align}
\widetilde b_2^2
=
4\lambda+3\,.
\end{align}
Consequently,
\begin{align}
\widetilde b_2^2
-
\left(\widetilde b_2^{\rm SU}\right)^2
=
1\,,
\label{eq:b-mismatch}
\end{align}
providing an independent low-order diagnostic of the departure from an
exact $\mathfrak{su}(1,1)$ chain.

\paragraph{Large-$n$ behavior and implications for the Krylov dynamics.}

Although the finite-$n$ Lanczos coefficients do not coincide with those of
an exact $\mathfrak{su}(1,1)$ representation, our numerical analysis at
large Krylov index $(n\gg1)$ indicates
\begin{align}
\widetilde a_n \sim 2n\,,
\qquad
\widetilde b_n \sim n\,,
\label{eq:asymptotic-dimensionless}
\end{align}
or, restoring physical units,
\begin{align}
a_n
\sim\frac{2n}{\beta}\,,
\qquad
b_n
\sim\frac{n}{\beta}\,.
\label{eq:asymptotic-physical}
\end{align}
The moment problem defined by the semiclassical partition function therefore
develops an \emph{asymptotically $\mathfrak{su}(1,1)$-like} structure, but
with nontrivial finite-$n$ corrections.

Importantly, the asymptotic slopes lie precisely at the critical point of
the effective $\mathfrak{su}(1,1)$ dynamics,
\begin{align}
\Gamma=2\kappa \,.
\label{eq:critical-su}
\end{align}
In the notation of Sec.~\eqref{sub42}, where
\begin{align}
a_n\simeq\gamma n\,,
\qquad
b_n\simeq\alpha\sqrt{c_1}\,n\,,
\end{align}
this condition becomes
\begin{align}
\gamma=2\alpha\sqrt{c_1}\,,
\end{align}
and therefore corresponds to the critical limit $\omega\to0$ of the
effective solution derived in Appendix~\eqref{appC}. The hyperbolic behavior
obtained by extrapolating a noncritical finite-range fit over the entire
Krylov chain should therefore not be interpreted as the true large-$n$
behavior of the moment problem defined by the semiclassical partition
function.

This observation also resolves the apparent mismatch in
\eqref{eq:survival-boundary}. The generating-function solution of Appendix~\eqref{appC} is obtained after replacing the full Lanczos
coefficients by their leading large-$n$ behavior. Extending this
approximation all the way to $n=0$ modifies precisely the region of the
Krylov chain that determines the survival amplitude. The resulting zeroth
amplitude is therefore not expected to reproduce the exact
$S_\beta(t)^*$ associated with the full Lanczos sequence.

\begin{figure}[t!]
\centering
\includegraphics[width=0.50\linewidth]{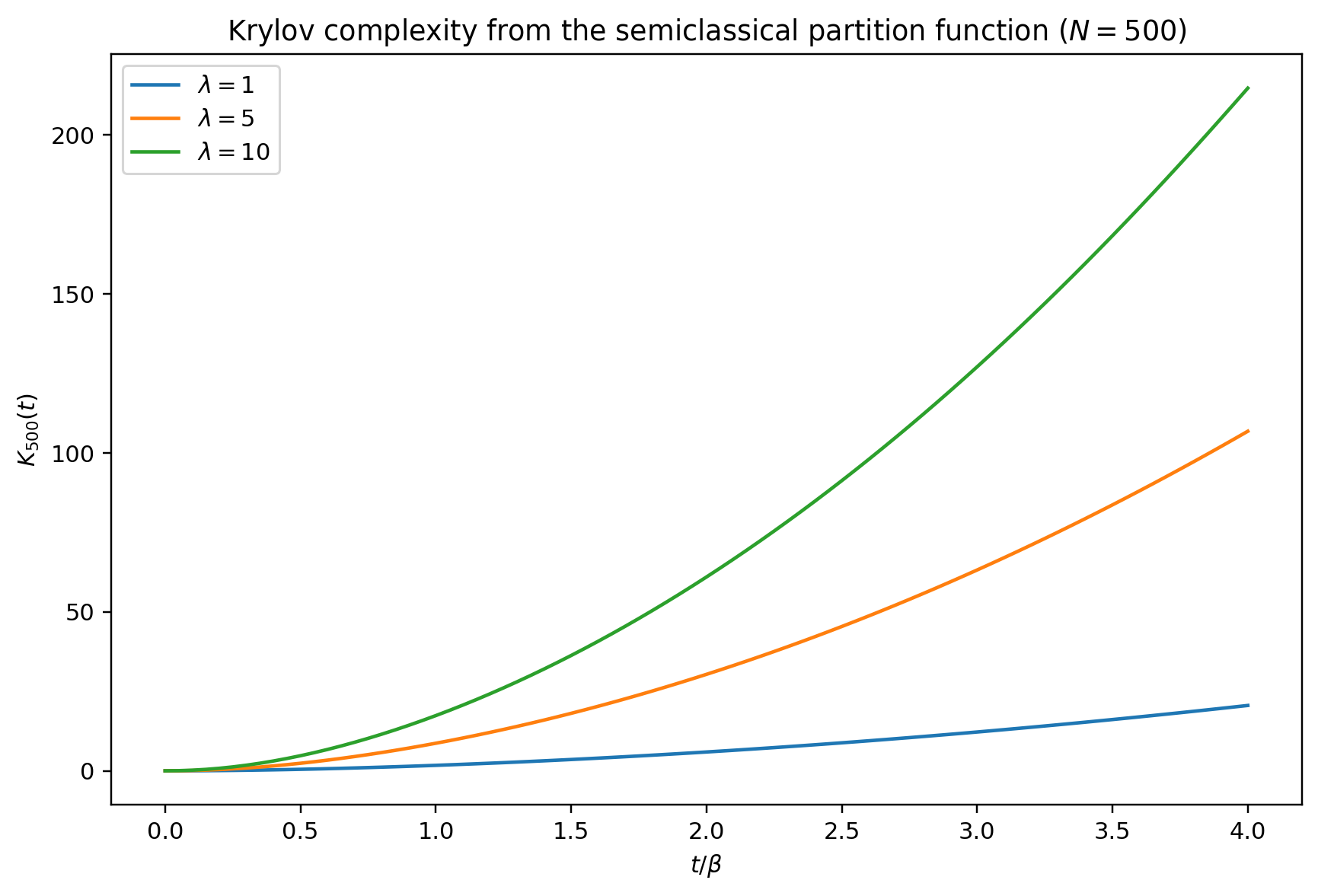}
\caption{Krylov spread complexity obtained by evolving the chain constructed
from the first $N=500$ Lanczos coefficients of the semiclassical moment
problem. The growth remains approximately quadratic over the accessible
time range and clearly departs from the exponential behavior obtained by
extrapolating a noncritical $\mathfrak{su}(1,1)$ approximation over the
full Krylov chain. Here again, $\lambda$ is defined in (\ref{E.3r}). }
\label{fig:7}
\end{figure}

The logically consistent construction is
\begin{align}
Z(\beta)
\longrightarrow
S_\beta(t)
\longrightarrow
\{\widetilde\mu_n\}
\longrightarrow
\{\widetilde a_n,\widetilde b_n\}
\longrightarrow
\{\psi_n(t)\}\,.
\label{eq:logic}
\end{align}
If each step is performed using the complete moment sequence and the full
Lanczos chain, the boundary condition
\begin{align}
\psi_0(t)=S_\beta(t)^*
\end{align}
is automatically satisfied.

For a fixed Hamiltonian and seed state, the Lanczos basis is fixed up to
trivial phase conventions. The failure of the asymptotic
$\mathfrak{su}(1,1)$ solution to reproduce the survival amplitude therefore
does not call for a different initial state or Lanczos basis, but rather
reflects the extrapolation of the asymptotic linear Lanczos coefficients
into the finite-$n$ regime. Consistently, evolving the Krylov chain using
the first $N=500$ Lanczos coefficients reconstructed from the semiclassical
moment problem shows a clear departure from the exponential behavior
predicted by the noncritical $\mathfrak{su}(1,1)$ approximation. As
illustrated in Fig.~(\ref{fig:7}), the spread complexity instead remains
approximately quadratic over the accessible time range, in agreement with the critical relation \eqref{eq:critical-su} and consistent with a persistent quadratic regime over the accessible time range.

\section{Krylov fluctuations: exact versus asymptotically
$\mathfrak{su}(1,1)$ dynamics}
\label{sec:krylov_fluctuations}

In Appendix~\eqref{appE}, we showed that the Lanczos chain generated by the
semiclassical BTZ partition function is not exactly
$\mathfrak{su}(1,1)$, although its coefficients become asymptotically
$\mathfrak{su}(1,1)$-like. Here, we provide a complementary diagnostic
based on fluctuations of the Krylov probability distribution.

Given a state expanded in the Krylov basis,
\begin{align}
|\psi(t)\rangle
=
\sum_{n=0}^{\infty}
\psi_n(t)|K_n\rangle\,,
\end{align}
we define
\begin{align}
P_n(t)
\equiv
|\psi_n(t)|^2\,.
\end{align}
The Krylov spread complexity is the mean position of the wavepacket along
the Krylov chain,
\begin{align}
\mathcal K_{\rm S}(t)
\equiv
\langle n\rangle
=
\sum_{n=0}^{\infty}
n\,P_n(t)\,.
\label{eq:K_definition}
\end{align}
Similarly, the second moment is \cite{Caputa:2021ori}
\begin{align}
\langle n^2\rangle
=
\sum_{n=0}^{\infty}
n^2 P_n(t)\,,
\label{eq:n2_definition}
\end{align}
and the variance is
\begin{align}
\operatorname{Var}[n](t)
=
\langle n^2\rangle
-
\mathcal K_{\rm S}(t)^2\,.
\label{eq:variance_definition}
\end{align}
A convenient measure of the relative width of the Krylov wavepacket is
the normalized variance
\begin{align}
\mathcal R(t)
\equiv
\frac{\operatorname{Var}[n](t)}
{\mathcal K_{\rm S}(t)^2}\,.
\label{eq:R_definition}
\end{align}
This quantity is the squared coefficient of variation of the Krylov
distribution. As we show below, its behavior provides a useful diagnostic
for distinguishing an exact $\mathfrak{su}(1,1)$ chain from the Lanczos
dynamics generated by the semiclassical partition function.

\subsection{Exact $\mathfrak{su}(1,1)$ dynamics}
\label{subsec:SL2_variance}

Consider the standard $\mathfrak{su}(1,1)$ Lanczos chain,
\begin{align}
a_n
&=
\gamma(n+k)+\delta\,,
&
b_n
&=
\alpha\sqrt{n(n+2k-1)}\,,
\label{eq:SL2_lanczos}
\end{align}
where $k$ labels the positive discrete-series representation.

Starting from the lowest-weight state,
\begin{align}
\psi_n(0)=\delta_{n0}\,,
\end{align}
the Krylov probability distribution takes the negative-binomial form
\begin{align}
P_n(t)
=
\frac{\Gamma(n+2k)}
{\Gamma(2k)n!}
\left[1-q(t)\right]^{2k}
q(t)^n\,.
\label{eq:negative_binomial}
\end{align}
The time dependence is entirely encoded in $q(t)$. In the hyperbolic
regime,
\begin{align}
\Omega^2
=
\alpha^2-\frac{\gamma^2}{4}
>0\,,
\end{align}
one has
\begin{align}
\frac{q(t)}{1-q(t)}
=
\frac{\alpha^2}{\Omega^2}
\sinh^2(\Omega t)\,.
\label{eq:q_SL2}
\end{align}

The generating function associated with
Eq.~\eqref{eq:negative_binomial} is
\begin{align}
\mathcal G(z,t)
&=
\sum_{n=0}^{\infty}
z^nP_n(t)
\nonumber\\
&=
\left(
\frac{1-q(t)}
{1-zq(t)}
\right)^{2k}.
\label{eq:SL2_generating}
\end{align}
The spread complexity follows directly,
\begin{align}
\mathcal K_{\rm S}(t)
&=
\left.
z\frac{\partial}{\partial z}
\mathcal G(z,t)
\right|_{z=1}
\nonumber\\
&=
2k\,\frac{q}{1-q}.
\label{eq:SL2_complexity}
\end{align}
The corresponding variance is
\begin{align}
\operatorname{Var}_{\mathfrak{su}(1,1)}[n]
=
\mathcal K_{\rm S}(t)
+
\frac{\mathcal K_{\rm S}(t)^2}{2k}\,,
\label{eq:SL2_variance}
\end{align}
and hence
\begin{align}
\mathcal R_{\mathfrak{su}(1,1)}(t)
=
\frac{1}{\mathcal K_{\rm S}(t)}
+
\frac{1}{2k}\,.
\label{eq:SL2_normalized_variance}
\end{align}
At sufficiently large complexity,
$\mathcal K_{\rm S}(t)\gg1$, the normalized variance therefore approaches
\begin{align}
\mathcal R_{\mathfrak{su}(1,1)}(t)
\longrightarrow
\frac{1}{2k}\,.
\label{eq:SL2_variance_asymptotic}
\end{align}
Thus, for an exact $\mathfrak{su}(1,1)$ chain, the normalized variance
approaches a constant fixed entirely by the representation parameter $k$.

A particularly simple example is $k=1/2$, for which
\begin{align}
b_n=\alpha n\,.
\end{align}
Equation~\eqref{eq:SL2_variance} then reduces to
\begin{align}
\operatorname{Var}_{\mathfrak{su}(1,1)}[n]
=
\mathcal K_{\rm S}(t)
+
\mathcal K_{\rm S}(t)^2\,,
\end{align}
so that
\begin{align}
\mathcal R_{\mathfrak{su}(1,1)}(t)
=
1+\frac{1}{\mathcal K_{\rm S}(t)}
\longrightarrow1\,.
\label{eq:khalf_variance}
\end{align}

\subsection{Effective representation parameter}
\label{subsec:keff}

Equation~\eqref{eq:SL2_variance} motivates the definition of an effective
representation parameter,
\begin{align}
k_{\rm eff}(t)
\equiv
\frac{
\mathcal K_{\rm S}(t)^2
}{
2\left[
\operatorname{Var}[n](t)-\mathcal K_{\rm S}(t)
\right]
}\,,
\label{eq:keff_definition}
\end{align}
whenever the denominator is nonzero. For exact
$\mathfrak{su}(1,1)$ dynamics,
Eq.~\eqref{eq:SL2_variance} immediately gives
\begin{align}
k_{\rm eff}(t)=k
\label{eq:keff_SL2}
\end{align}
at all times. A time-dependent $k_{\rm eff}$ therefore provides a direct
diagnostic of departures from an exact $\mathfrak{su}(1,1)$ structure.

\begin{figure}[t!]
    \centering
    \includegraphics[width=0.55\linewidth]{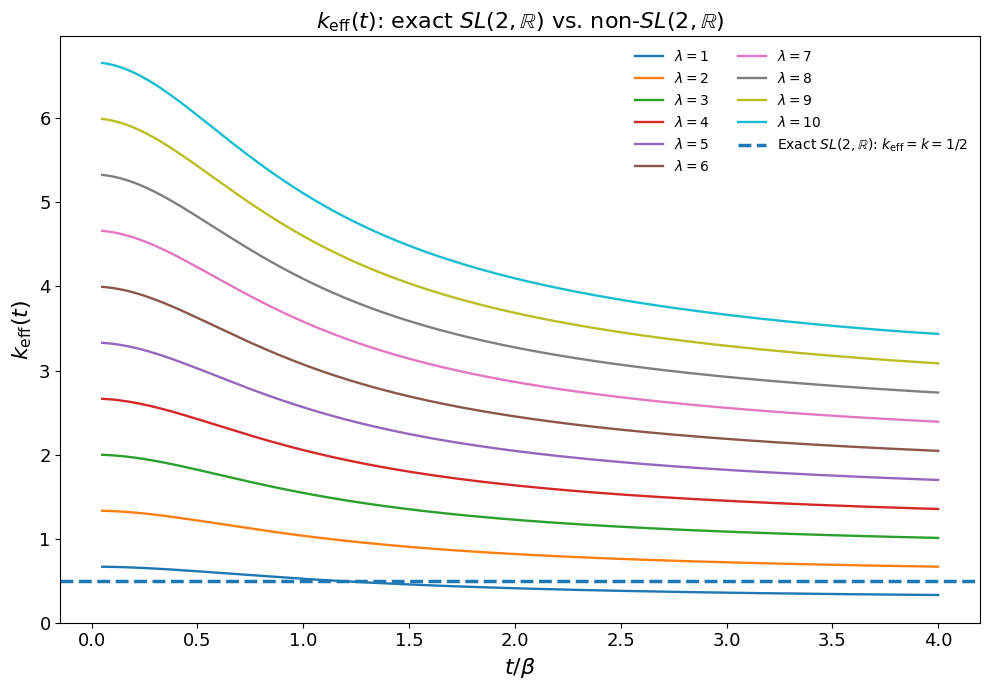}
    \caption{Effective representation parameter $k_{\rm eff}$ as a function
    of the dimensionless time $\tau=t/\beta$ for the Lanczos chain generated
    by the semiclassical moment problem, shown for several values of
    $\lambda$, defined in Eq.~\eqref{E.3r}. The dashed horizontal line corresponds to the constant
    $k=1/2$ of an exact $\mathfrak{su}(1,1)$ reference chain. The time
    dependence of $k_{\rm eff}$ provides a direct diagnostic of the departure
    from exact $\mathfrak{su}(1,1)$ dynamics.}
    \label{fig:8}
\end{figure}

For the Lanczos chain generated by the semiclassical moment problem,
$k_{\rm eff}$ varies with time, as illustrated in
Fig.~\eqref{fig:8}. This provides an independent fluctuation-based
manifestation of the finite-$n$ corrections identified in
Appendix~\eqref{appE}.

\subsection{Analytic short-time fluctuations}
\label{subsec:short_time_variance}

The short-time behavior of these fluctuations can also be characterized
analytically. Since the moment problem in Appendix~\eqref{appE} is formulated
in terms of the dimensionless Hamiltonian $x=\beta H$, it is convenient to
introduce the corresponding dimensionless time
\begin{align}
\tau\equiv\frac{t}{\beta}\,.
\end{align}
Using the first Lanczos coefficients obtained from the semiclassical
moments, the leading Krylov probabilities are
\begin{align}
|\psi_0(\tau)|^2
&=
1-2\lambda\tau^2
+2\lambda(\lambda+1)\tau^4
+\mathcal O(\tau^6)\,,
\nonumber\\
|\psi_1(\tau)|^2
&=
2\lambda\tau^2
-
\left(
4\lambda^2+\frac{7}{2}\lambda
\right)\tau^4
+\mathcal O(\tau^6)\,,
\nonumber\\
|\psi_2(\tau)|^2
&=
\frac{\lambda(4\lambda+3)}{2}\tau^4
+\mathcal O(\tau^6)\,.
\label{eq:phi2_short}
\end{align}
Here $\lambda$ is defined in Eq.~\eqref{E.3r}. It follows that the spread
complexity behaves as
\begin{align}
\mathcal K_{\rm S}(\tau)
=
2\lambda\tau^2
-\frac{\lambda}{2}\tau^4
+\mathcal O(\tau^6)\,.
\label{eq:K_short_nonSL2}
\end{align}
The second moment is
\begin{align}
\langle n^2\rangle
=
2\lambda\tau^2
+
\left(
4\lambda^2+\frac{5}{2}\lambda
\right)\tau^4
+
\mathcal O(\tau^6)\,,
\label{eq:n2_short_nonSL2}
\end{align}
and therefore
\begin{align}
\operatorname{Var}[n]
=
2\lambda\tau^2
+
\frac{5}{2}\lambda\tau^4
+
\mathcal O(\tau^6)\,.
\label{eq:variance_short_nonSL2}
\end{align}

\noindent
The normalized variance consequently has the expansion
\begin{align}
\mathcal R(\tau)
=
\frac{1}{2\lambda\tau^2}
+
\frac{7}{8\lambda}
+
\mathcal O(\tau^2)\,.
\label{eq:R_short_nonSL2}
\end{align}
The divergence as $\tau\to0$ is purely kinematical: both the mean and the
variance begin at order $\tau^2$, whereas
$\mathcal K_{\rm S}^2$ starts at order $\tau^4$.

Using Eq.~\eqref{eq:keff_definition}, we also find
\begin{align}
\operatorname{Var}[n]
-
\mathcal K_{\rm S}
=
3\lambda\tau^4
+
\mathcal O(\tau^6)\,,
\end{align}
and hence
\begin{align}
k_{\rm eff}(\tau\to0)
=
\frac{2\lambda}{3}
+
\mathcal O(\tau^2)\,.
\label{eq:keff_short}
\end{align}
Thus, the short-time expansion assigns a well-defined local effective
representation parameter to the semiclassical Krylov chain. For an exact
$\mathfrak{su}(1,1)$ representation this parameter would remain constant
throughout the evolution, whereas the numerical results in
Fig.~\eqref{fig:8} show that $k_{\rm eff}$ evolves with time. The fluctuation
analysis therefore provides a complementary diagnostic of the departure
from exact $\mathfrak{su}(1,1)$ dynamics.

\bibliography{refK}
\bibliographystyle{utphysmodb}
\end{document}